\documentclass[sigplan,screen]{acmart}

\DeclareMathAlphabet{\mathcal}{OMS}{cmsy}{m}{n}

\usepackage{amsmath,amssymb,amsfonts}
\usepackage{algorithmic}
\usepackage{graphicx}
\usepackage{booktabs}
\usepackage{colortbl}
\usepackage{multirow}
\usepackage{textcomp}
\usepackage{pifont}
\usepackage{autobreak}
\usepackage{amssymb}
\usepackage{xcolor}
\usepackage{soul}
\soulregister\cite7 % 针对\cite命令
\soulregister\underline7 % 针对\cite命令
\soulregister\ding7 % 针对\cite命令
\soulregister\citep7 % 针对\citep命令
\soulregister\citet7 % 针对\citet命令
\soulregister\ref7 % 针对\ref命令
\soulregister\pageref7 % 针对\pageref命令
\usepackage[ruled,linesnumbered]{algorithm2e}
\usepackage{footnote}
\usepackage{enumitem}
\usepackage{diagbox}
\usepackage{longtable}
\renewcommand{\hl}{}
\AtBeginDocument{%
  }

\copyrightyear{2024} 
\acmYear{2024} 
\setcopyright{rightsretained} 
\acmConference[ASPLOS '24]{29th ACM International Conference on Architectural Support for Programming Languages and Operating Systems, Volume 1}{April 27-May 1, 2024}{La Jolla, CA, USA}
\acmBooktitle{29th ACM International Conference on Architectural Support for Programming Languages and Operating Systems, Volume 1 (ASPLOS '24), April 27-May 1, 2024, La Jolla, CA, USA}
\acmDOI{10.1145/3617232.3624865}
\acmISBN{979-8-4007-0372-0/24/04}

\begin{document}

%%
%% The "title" command has an optional parameter,
%% allowing the author to define a "short title" to be used in page headers.
\title[Cocco: Hardware-Mapping Co-Exploration towards Memory ...]{Cocco: Hardware-Mapping Co-Exploration towards Memory Capacity-Communication Optimization}

%%
%% The "author" command and its associated commands are used to define
%% the authors and their affiliations.
%% Of note is the shared affiliation of the first two authors, and the
%% "authornote" and "authornotemark" commands
%% used to denote shared contribution to the research.
\author{Zhanhong Tan}
\email{tanzh19@mails.tsinghua.edu.cn}
\affiliation{%
  \institution{IIIS, Tsinghua University}
  \city{Beijing}
  \country{China}
}

\author{Zijian Zhu}
\email{zhuzj23@mails.tsinghua.edu.cn}
\affiliation{%
  \institution{IIIS, Tsinghua University}
  \city{Beijing}
  \country{China}
}

\author{Kaisheng Ma}
\authornote{Corresponding author.}
\email{kaisheng@mail.tsinghua.edu.cn}
\affiliation{%
  \institution{IIIS, Tsinghua University}
  \city{Beijing}
  \country{China}
}

%%
%% By default, the full list of authors will be used in the page
%% headers. Often, this list is too long, and will overlap
%% other information printed in the page headers. This command allows
%% the author to define a more concise list
%% of authors' names for this purpose.
% \renewcommand{\shortauthors}{Trovato et al.}

%%
%% The abstract is a short summary of the work to be presented in the
%% article.
\begin{abstract}
    Memory is a critical design consideration in current data-intensive DNN accelerators, as it profoundly determines energy consumption, bandwidth requirements, and area costs.
    As DNN structures become more complex, a larger on-chip memory capacity is required to reduce data movement overhead, but at the expense of silicon costs.
    Some previous works have proposed memory-oriented optimizations, such as different data reuse and layer fusion schemes.
    However, these methods are not general and potent enough to cope with various graph structures. 

    In this paper, we explore the intrinsic connection between network structures and memory features to optimize both hardware and mapping.
    First, we introduce a graph-level execution scheme with a corresponding dataflow and memory management method.
    This scheme enables the execution of arbitrary graph patterns with high data reuse and low hardware overhead.
    Subsequently, we propose Cocco, a hardware-mapping co-exploration framework leveraging graph-level features of networks.
    It aims to minimize communication overhead, such as energy consumption and bandwidth requirements, with a smaller memory capacity.
    We formulate the graph-partition scheduling and memory configuration search as an optimization problem and employ a genetic-based method to achieve efficient co-exploration for large and irregular networks.
    Experiments demonstrate that Cocco obtains lower external memory access, lower bandwidth requirements, and more stable optimization for graph partition compared to the greedy algorithm and dynamic programming introduced in prior works. 
    Cocco also reduces the costs by 1.89\% to 50.33\% using co-exploration compared to other typical methods.
\end{abstract}

%%
%% The code below is generated by the tool at http://dl.acm.org/ccs.cfm.
%% Please copy and paste the code instead of the example below.
%%


%%
%% Keywords. The author(s) should pick words that accurately describe
%% the work being presented. Separate the keywords with commas.

%% A "teaser" image appears between the author and affiliation
%% information and the body of the document, and typically spans the
%% page.
% \begin{teaserfigure}
%   \includegraphics[width=\textwidth]{sampleteaser}
%   \caption{Seattle Mariners at Spring Training, 2010.}
%   \Description{Enjoying the baseball game from the third-base
%   seats. Ichiro Suzuki preparing to bat.}
%   \label{fig:teaser}
% \end{teaserfigure}

% \received{20 February 2007}
% \received[revised]{12 March 2009}
% \received[accepted]{5 June 2009}

%%
%% This command processes the author and affiliation and title
%% information and builds the first part of the formatted document.
\maketitle

\section{Introduction}
\label{sec:intro}
The evolution of neural network topology has driven the remarkable progress of artificial intelligence from the early single-layer perceptron~(SLP)~\cite{slp1, slp2} and multi-layer perceptron~(MLP)~\cite{mlp1, mlp2, mlp3} to modern DNNs with plain~\cite{alexNet, vgg}/inception~\cite{inception}/residual~\cite{res, mobilenet} structures based on manual design, and even irregular structures using neural architecture search~(NAS)~\cite{nas1,nasnet} or random network generation~\cite{random-gen}.
These technological innovations have resulted in increasingly complex computation graphs, which pose challenges for efficient memory design and deployment.

\begin{figure}[!t]
    % \vspace{-4mm}
    \centering
    \includegraphics[width=\columnwidth]{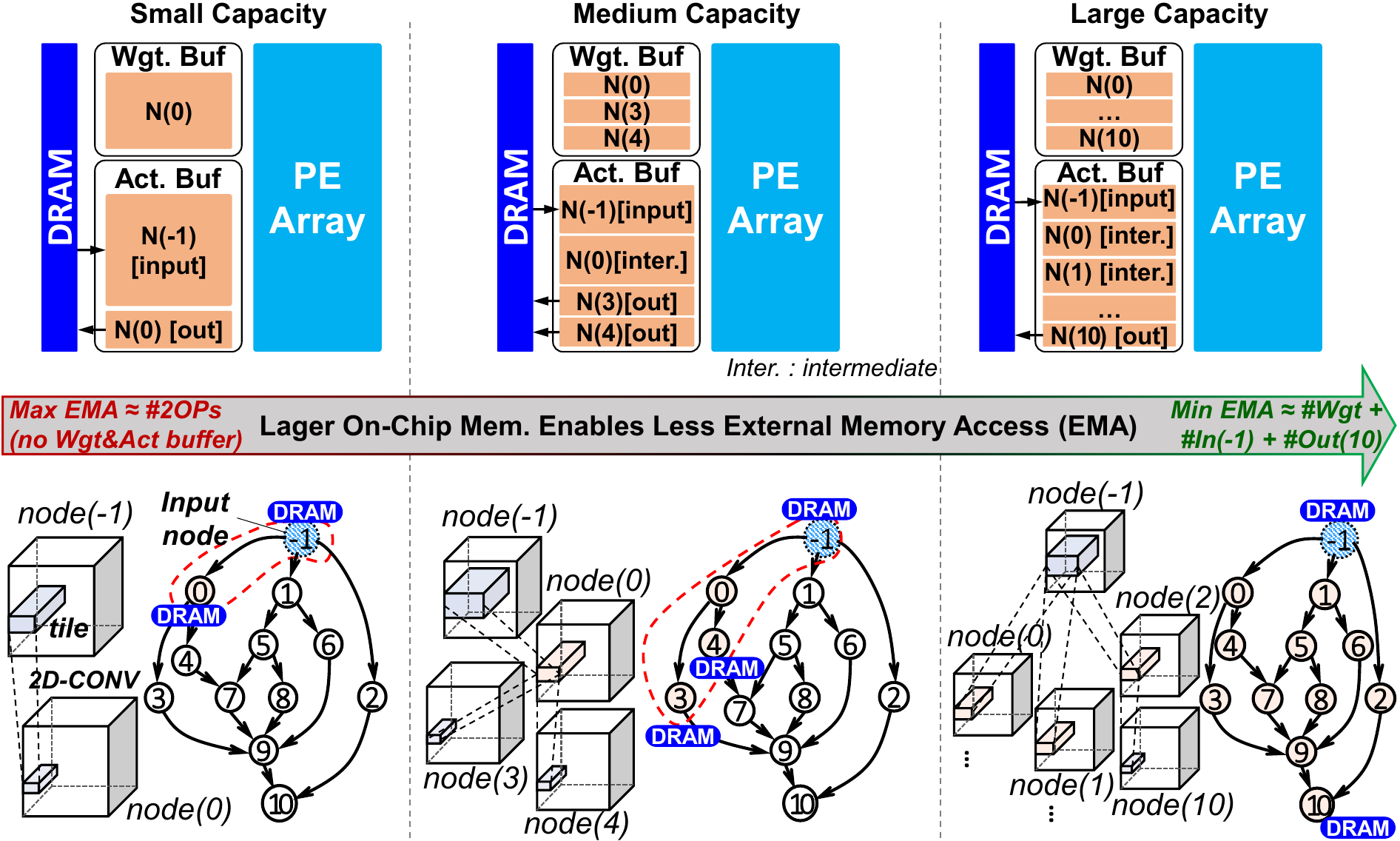}
    \caption{The effect of different memory capacities for a computation graph. Intermediate results can be buffered in the on-chip memory if it is large enough. The on-chip memory of small capacity can only buffer two nodes (marked in the red dotted box), and the larger memory can cover a larger subgraph (right side).}
    \label{fig:eval}
    \vspace{-4mm}
\end{figure}

Memory design is crucial in the accelerator system, as it performs data preparation at the start of each processing stage according to the scheduling scheme, determining energy consumption, bandwidth requirements, and area costs.
Figure~\ref{fig:eval} shows the trade-off between the on-chip memory size and the external memory access in DNN accelerators.
\hl{A smaller on-chip buffer (left side) saves area but requires more data reloading.
A larger buffer (right side) can reduce external memory access and save energy and bandwidth but at the cost of increasing the memory overhead.}
An excessively large SRAM may not be feasible due to the high silicon area cost, typically ranging from 1 to 2~mm$^2$/MB in 12nm, and the high energy overhead, dozens of times that of a MAC operation for a large SRAM.

Therefore, the \textbf{key problem} is: \textit{between the two extremes in Figure~\ref{fig:eval}, how to find an appropriate \ul{memory configuration} with \ul{efficient workload mapping and data management}, especially under the growing complexity of neural network architectures}.

The critical status of memory design has attracted extensive research.
Most previous studies focus on simple layer-level optimization (the left one of Figure~\ref{fig:eval}) by applying loop transformation techniques such as tiling and reordering to fit the memory size and reuse the on-chip data~\cite{nnbaton, dataflow2, basic-dataflow1, interstellar, CoSA}.
In addition, \hl{several} works also guide the memory capacity and hierarchy design using design-space exploration~\cite{digamma, understand, dse, hasco, lower-bound}.
However, these layer-level optimizations are confined to the limited intra-layer reuse, which is insufficient for memory-intensive networks.
A subgraph-level scheme (e.g., the middle one and the right one of Figure~\ref{fig:eval}) provides a larger optimization space via inter-layer reuse~\cite{fused-CNN,fused-cnn2,irregular,chaos} to reduce the I/O overhead.
Therefore, this paper aims to \textit{leverage the subgraph-level computing flow to optimize the memory capacity and external communication for networks with any topology}.

However, there are \textbf{three primary challenges} to fully exploit the subgraph-level optimization.

First, \textit{we need a general execution flow for any sub-graph.}
Due to the various kernel sizes and strides, a parent node in a subgraph may have unbalanced data requirements from its consumers, which makes it difficult to determine the tensor tiling scheme and the memory allocation for each node (layer).
In the traditional single-layer execution, we usually divide a large tensor into loop tiles, which are processed through a series of regular computing steps. 
Similarly, we want the sub-graph execution to be a series of elementary computing steps with a simple control flow.

\hl{Second, \textit{we require a suitable memory management method for the subgraph execution}.
Due to complicated dependency among nodes in a subgraph, careful management is needed to reuse overlapping and inter-layer intermediate data.}

Solving these two challenges contributes to a basic hardware execution model compatible with subgraph-level optimization.
However, we also encounter the third challenge: \textit{how to partition a model into subgraphs and how much memory to allocate.}
The optimization space is huge, so we need to devise a search method with high sampling efficiency to find a proper subgraph partition and memory configuration result.

In this paper, we first introduce a complete graph-level scheme for memory.
In particular, it contains a consumption-centric flow that enables the execution of arbitrary subgraphs with low memory footprints (\textit{for challenge 1}).
Accordingly, we provide an explicit memory dataflow and the corresponding memory management scheme for effective data reuse (\textit{for challenge 2}).
Building on the graph-level memory scheme, we propose Cocco, a hardware-mapping co-exploration framework, to establish a connection between model features and the memory configuration (\textit{for challenge 3}).

Cocco aims to find a combination of on-chip buffers and the corresponding graph-level scheduling for lower memory and communication overhead. 
In particular, we develop a genetic-based algorithm to efficiently explore the search space of graph partitions and the associated memory configuration for a series of neural networks.

In summary, this work makes the following contributions:
\begin{itemize}
    \item \textbf{Subgraph execution scheme.} 
    We first introduce a consumption-centric flow to determine a low-cost execution sequence by throttling and aligning the dataflow.
    \item \hl{\textbf{Efficient dataflow and memory management} for subgraph data reuse.} We propose a memory management scheme featuring multiple reconfigurable regions and the corresponding dataflow to support arbitrary subgraph execution with full data reuse.
    \item \textbf{Hardware-mapping co-exploration framework.} Based on the subgraph execution scheme and memory dataflow, we propose Cocco, a genetic-based framework combining the graph-level partition and memory design-space exploration together.
    Cocco achieves 1.89\% to 50.33\% lower costs (lower communication with a smaller size) using co-exploration in contrast to other methods.
\end{itemize}

\section{Background and Motivation}
\subsection{Design of Neural Network Accelerators}
\hl{The DNN accelerator unit is the most basic execution unit in a computing system}, on top of which, we can scale it out to many-core, many-socket, and many-drawer systems~\cite{Ascend1,TPUv2,dojo, Wormhole}.
An accelerator unit usually employs a processing element (PE) array on a sophisticated interconnection network to enable efficient tensor-level computation.
Each PE typically contains local scratchpads and ALUs to process basic data packets.
The global buffer and the weight buffer store activations and weights, and they are generally located next to the PE array to serve as the data interface and manage data between the PE array and the external memory (e.g., DRAM or other cores).
Due to the limited capacity of the global buffer, \hl{the compiler has to partition the network execution into a series of elementary workloads} that are scheduled along the parallel spatial resources and the temporal dimension~\cite{nnbaton, tangram, atomic}.
The capacity of the global buffer usually dominates the external memory access and bandwidth requirements, significantly impacting system performance.
If the global memory is larger, it is more likely to buffer more intermediate data and avoid data being evicted to DRAM.
As shown in Figure~\ref{fig:eval}, a larger buffer expands the scope of elementary workloads from a single layer to a larger subgraph, reducing the communication overhead.

\begin{figure}[!t]
    % \vspace{-4mm}
    \centering
    \includegraphics[width=\columnwidth]{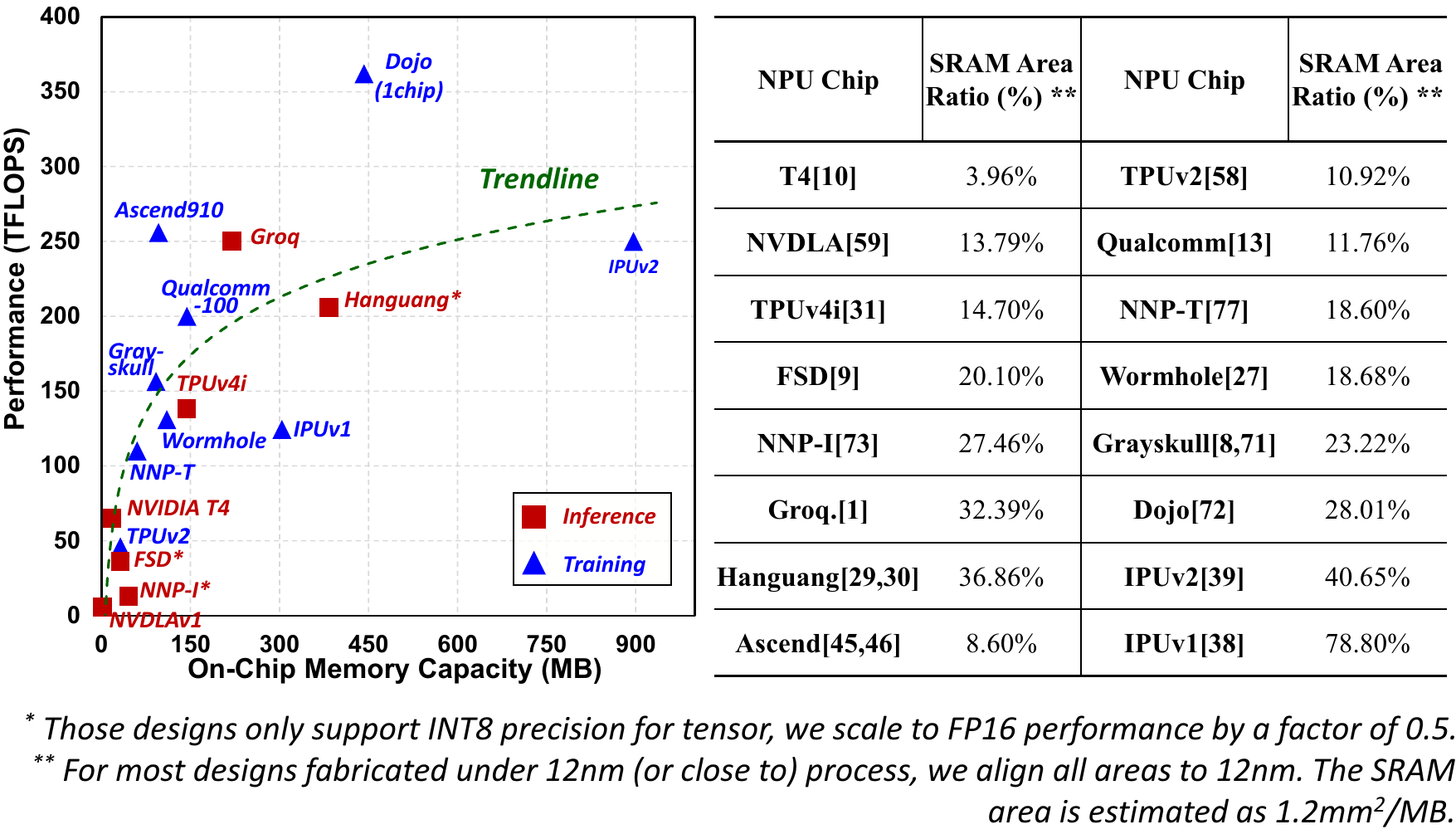}
    \caption{Left: performance v.s. memory capacity of several industrial NPUs. Right: a summary of SRAM area ratio in these accelerators.}
    \label{fig:survey}
    % \vspace{-5mm}
\end{figure}

However, choosing an appropriate memory specification is always a challenge. 
In Figure~\ref{fig:survey}, we surveyed 16 popular industrial neural network processors with various memory/performance/area characteristics, where nine of them target the training domain~\cite{Ascend1,Ascend2,Grayskull1,Grayskull2,IPUv1,IPUv2,NNP-T,Wormhole,TPUv2,Qualcomm, dojo} and seven target model  inference~\cite{FSD,Hanguang1,Hanguang2,NNP-I,TPUv4i,T4,nvdla,Groq}.
According to the survey, we can observe several trends as follows:
\begin{enumerate}
    \item Memory occupies a significant portion of the silicon footprint on an NPU chip, ranging from 4\% to 79\% of the area, with capacities from 2.5MB to 896MB.
    \item Figure~\ref{fig:survey} Left shows a trend of diminishing marginal benefit of memory capacity. This is because there is a critical capacity to meet the data reuse and bandwidth requirement at the beginning, and the increments become negligible with higher memory capacity.
    \item We can infer that there is a saturated capacity equivalent to the ideal unlimited memory, especially for the inference design.
    For example, Hanguang~\cite{Hanguang1} is a special SRAM-only inference system without DDR, and the 394MB buffers are large enough to hold the intermediate data in their scenarios.
\end{enumerate}

This survey implies a design trade-off between memory capacity and performance based on workloads and commercial considerations.
Motivated by the observations above, this paper aims to provide several memory design considerations and study the connection between workload features and memory capacity in an NPU accelerator.

\subsection{Workload Deployment}

A neural network is usually executed in a DNN accelerator with layer or graph granularities based on the buffer capacity and dataflow.
\subsubsection{Layer-level Assignment}
This manner assigns tasks layer by layer.
Most previous studies employ a tiling-based layer-wise execution manner~\cite{nnbaton, understand, timeloop, mindmap, GAMMA, marvel}, which elaborates the tiling sizes of tensors to fit in the accelerator buffers and maintain performance.
A proper tiling scheme should overlap the data loading latency with the computing time of each tile and try to reduce the repeated access of local weight buffers.
Tiles of data are transferred between the external memory and the global buffer, and PEs subsequently fetch data from the global to their local buffers.
Given the larger bit-width of partial sums (e.g., 24bit partial sums v.s. 8bit inputs in Simba), the output-centric tiling scheme is more commonly used to calculate the final results before writing back to the global buffer~\cite{nnbaton}.

\begin{figure}[!t]
    % \vspace{-3mm}
    \centering
    \includegraphics[width=\columnwidth]{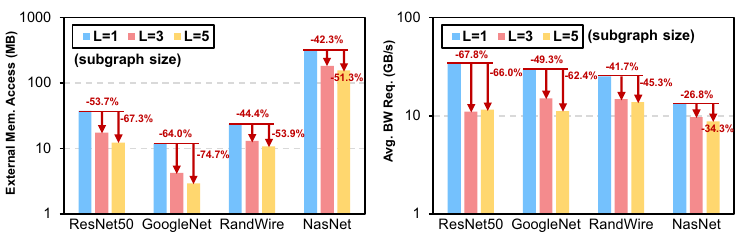}
    % \vspace{-3mm}
    % \setlength{\abovecaptionskip}{-4mm}
    \caption{Evaluations on subgraphs fusing different number of layers (denoted as \textit{L}=1,3,5).
    Y-axis is in the log domain.
    The 2TOPS NPU accelerator is configured with a 1MB global buffer and a 1.125MB weight buffer.
    The bandwidth requirement of weights is from the prefetch of the next subgraph, while that of activations is from the inputs and outputs of each subgraph.}
    \label{fig:bg_exp}
    % \vspace{-2mm}
\end{figure}

\subsubsection{Graph-level Assignment}

Unlike the layer-level assignment that restrains from leveraging inter-layer reuse, a graph-level assignment processes several layers of a neural network as a whole.
To demonstrate the effectiveness of the layer-level assignment, we evaluate four networks on a 2TOPS accelerator model, as shown in Figure~\ref{fig:bg_exp}.
The results show that fusing layers into subgraphs significantly reduces external memory access by 42.3\% $\sim$ 74.7\% and average bandwidth requirements by 26.8\% $\sim$ 67.8\%.
However, the improvements of larger subgraphs are marginal, indicating that there is an optimal trade-off between inter-layer reuse and subgraph size, which determines the memory requirement.
For example, executing three-layer subgraphs reduces external memory access by 53.7\% in ResNet50, while executing five-layer subgraphs only further reduces it by 13.6\%.

Several works have studied inter-layer reuse and graph partition.
However, they have several limitations in terms of performance and flexibility.
LCP~\cite{lcp} groups similar layers into a cluster and executes them as a whole, which makes it challenging to generalize into an arbitrary graph.
Fused-CNN~\cite{fused-CNN} and SR-CNN~\cite{fused-cnn2} fuse large contiguous layers for plain networks using manually-designed strategies.
Irregular-NN~\cite{irregular} attempts to execute a complex subgraph using a DP-based algorithm, but the constrained search space limits the exploration.

To overcome these challenges, we propose an end-to-end framework that automatically optimizes the graph partition and memory configuration for any neural network.
Our framework consists of two main components: a graph-level dataflow and a hardware-mapping co-exploration algorithm.
We first introduce the graph-level dataflow and its hardware implementation.
Then, we present Cocco, an efficient algorithm that explores the trade-offs among memory configurations and graph partition schemes based on workload features.
\section{The Proposed Graph-Level Scheme}
\label{sec:memory}

To execute layers on an NPU core in a graph-level manner, we need an effective approach to reuse intermediate data and decide the memory allocation.
This section presents our comprehensive scheme for subgraph execution, which addresses the first two challenges mentioned in Section~\ref{sec:intro}.
First, we describe a multi-layer execution flow that minimizes the memory footprint by a friendly tiling approach (\textit{for challenge~1}).
Second, we explain how to implement this flow on a real NPU using an efficient data reuse pattern (\textit{for challenge~2}).
The consistent target is to reduce the memory footprint and be friendly to implementation.

\subsection{Subgraph execution scheme}
\label{sec:exe}

It is common practice for the layer-level scheduling to partition the output tensor into several tiles as layer-level elementary operations~\cite{nnbaton, atomic, simba-vlsi, simba}, simplifying the scheduling and instruction generation.
Likewise, our high-level idea is also to generate a series of explicit \textbf{subgraph-level elementary operations}.
However, we need to address the challenges of various kernel sizes and strides in different paths to prevent unbalanced data production and unnecessary memory.

\begin{figure}[!b]
    % \vspace{-4mm}
    \centering
    \includegraphics[width=\columnwidth]{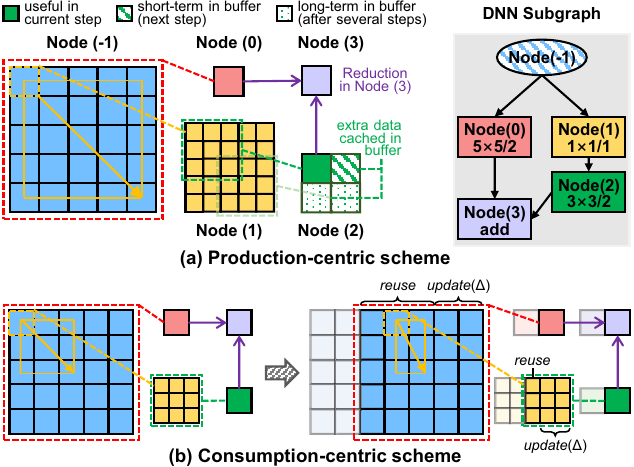}
    \caption{A conceptual comparison between two manners to process a subgraph. The node marked with a negative number represents the input node. \hl{The corresponding subgraph is shown in the upper right, where \underline{$F\times F/s$} refers to the convolution kernel size ($F$) and stride ($s$).}}
    \label{fig:atomic-step}
    % \vspace{-2mm}
\end{figure}

A model's subgraph consists of multiple layers (nodes) with dependencies. Section~\ref{sec:cocco} provides detailed information on subgraph partition.
In Figure~\ref{fig:atomic-step}(a), we present a straightforward {\textbf{production-centric scheme}} for executing a subgraph with different kernel sizes in two branches, deriving tile sizes of the subsequent layers based on the predetermined input tile sizes.
For example, we can produce a $1\times1$ tile of Node\textsf{(0)} and a $2\times2$ tile of Node\textsf{(2)} with a given $5\times5$ feature map of input Node\textsf{(-1)}.
In this case, these intermediate results only reduce to $1\times1$ in Node\textsf{(3)}, limited by the smallest input of Node\textsf{(0)}, so the remaining results of Node\textsf{(2)} can not be consumed immediately.
\hl{As shown in Figure~\ref{fig:atomic-step}, three extra data of Node\textsf{(2)} along with sixteen extra source data of Node\textsf{(1)} take up extra memory space.}
There are more redundant cached data when the subgraph becomes larger and more complicated.
Disadvantages of this manner are attributed to the production-centric idea that consumes all related activations from the producers at once.

To avoid the memory overhead of storing unused data, we propose a \textbf{consumption-centric scheme} in Figure~\ref{fig:atomic-step}(b), where results of each node are \textit{produced on demand based on consumer(s)} (i.e., output node(s)).
For example, given a $1\times1$ tile of Node\textsf{(3)}, we derive the $1\times1$ tile size for Node\textsf{(2)}, which subsequently decides a $3\times3$ tile for Node\textsf{(1)}.

\begin{figure}[!t]
    % \vspace{-2mm}
    \centering
    \includegraphics[width=\columnwidth]{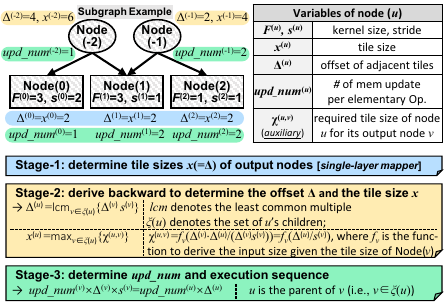}
    \caption{The flow to determine the execution scheme of a subgraph (i.e., the computed tile size of each node, the tile offset, and the processing sequence of nodes).
    For simplicity, we discuss the 1D-CONV in this example and it is similar in the 2D-CONV case.}
    \label{fig:atomic-desc}
    % \vspace{-2mm}
\end{figure}

The backward-derivation for each producer node is non-trivial because of diverse kernel sizes and strides in different paths.
Therefore, we propose a three-stage flow to determine the behavior of each node, as illustrated in Figure~\ref{fig:atomic-desc}.
The high-level idea is to let output nodes drive the whole execution and match the data consumption and production in each subgraph-level elementary operation.

\textbf{The stage-1} is similar to the traditional single-layer scheduling, where the tile size is optimized for higher computation utilization.
In order to hold a larger subgraph, the tile size tends to be smaller. In the 1D-CONV example, we set the tile size to be $2$ for output nodes.

\textbf{The stage-2} aims to determine the data update offset $\Delta$ and the memory allocation size $x$ for each node based on the consumer(s), processing in the reverse topological order.
We use the least common multiply (LCM) operation to determine $\Delta^{(u)}$ of producers for aligning different input offset requirements ($\Delta^{(v)}s^{(v)}$) from consumers.
Hence, one producer update may correspond to multiple updates of a consumer.
For example, $\Delta^{(-2)}=\text{lcm}\{\Delta^{(0)}s^{(0)},\Delta^{(1)}s^{(1)}\}=4=2\Delta^{(1)}s^{(1)}$, one update of Node\textsf{(-2)} corresponds to two updates of Node\textsf{(1)}.
As for the tile size deduction, $f_v(\Delta^{(u)}/s^{(v)})$ is to derive the required input tile size $\chi^{(u,v)}$ for output node $v$\footnote{For example, assume node $v$ is a convolution layer with kernel size $F^{(v)}$ and stride $s^{(v)}$, then $f_v(x) = F^{(v)}+(x-1)\times s^{(v)}$.}, where $\Delta^{(u)}/s^{(v)}$ is the consumer offset (updated data) per producer $u$ update.
The maximum result $\chi^{(u,v)}$ of all outputs $v$ is the tile size $x^{(u)}$ of input node $u$. In this example, $x^{(-2)}=\max\{f_0(2), f_1(4)\}=6$ and $x^{(-1)}=\max\{f_1(2), f_2(2)\}=4$.

\begin{figure}[!t]
    % \vspace{2mm}
    \centering
    \includegraphics[width=\columnwidth]{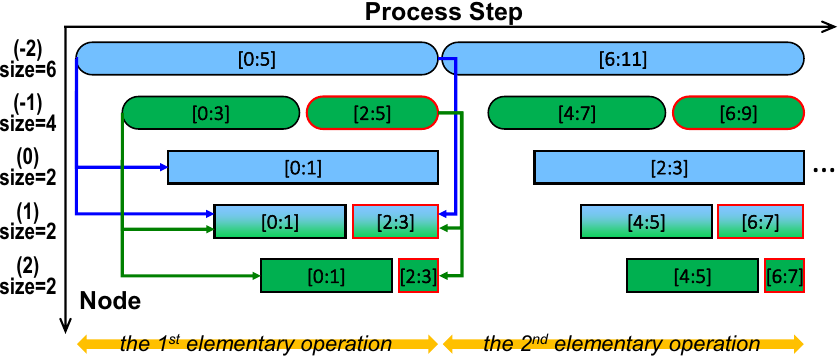}
    \caption{The memory snapshot during two subgraph elementary operations based on the execution scheme of Figure~\ref{fig:atomic-desc} example. The allocated memory size and update offset correspond to $x$ and $\Delta$, respectively (the [$m$:$n$] notation denotes data ranging from index $m$ to $n$).
    The arrows denote the data dependency according to the node relation in the subgraph.}
    \label{fig:atomic-flow}
    % \vspace{-2mm}
\end{figure}

As mentioned above, since we use LCM to align production and consumption, one producer update may correspond to multiple updates of a consumer.
In \textbf{the stage-3}, we use $upd\_num$ to represent the number of memory update per subgraph elementary operation.
The generated result of the example in Figure~\ref{fig:atomic-desc} is shown in Figure~\ref{fig:atomic-flow}.
$upd\_num$ of Node\textsf{(-1)}, Node\textsf{(1)}, and Node\textsf{(2)} are two, where the second updates are highlighted in red boxes.
Note that the $ \{ upd\_num^{(-2)}, \dots, upd\_num^{(2)} \} $
solution is not unique, but the unique co-prime one $\{ 1,2,1,2,2 \}$ corresponds to the minimal elementary operation.

\hl{The proposed flow is based on a general directed acyclic computation graph and is not limited to specific layer features.
% can apply to any computation graph with a directed acyclic graph (DAG) topology.
In this way, we can determine the execution scheme for any complex irregular network like NasNet~\cite{nasnet} and RandWire~\cite{random-gen}.}

\subsection{Memory Management for the subgraph execution}

\begin{figure}[!b]
    % \vspace{2mm}
    \centering
    \includegraphics[width=\columnwidth]{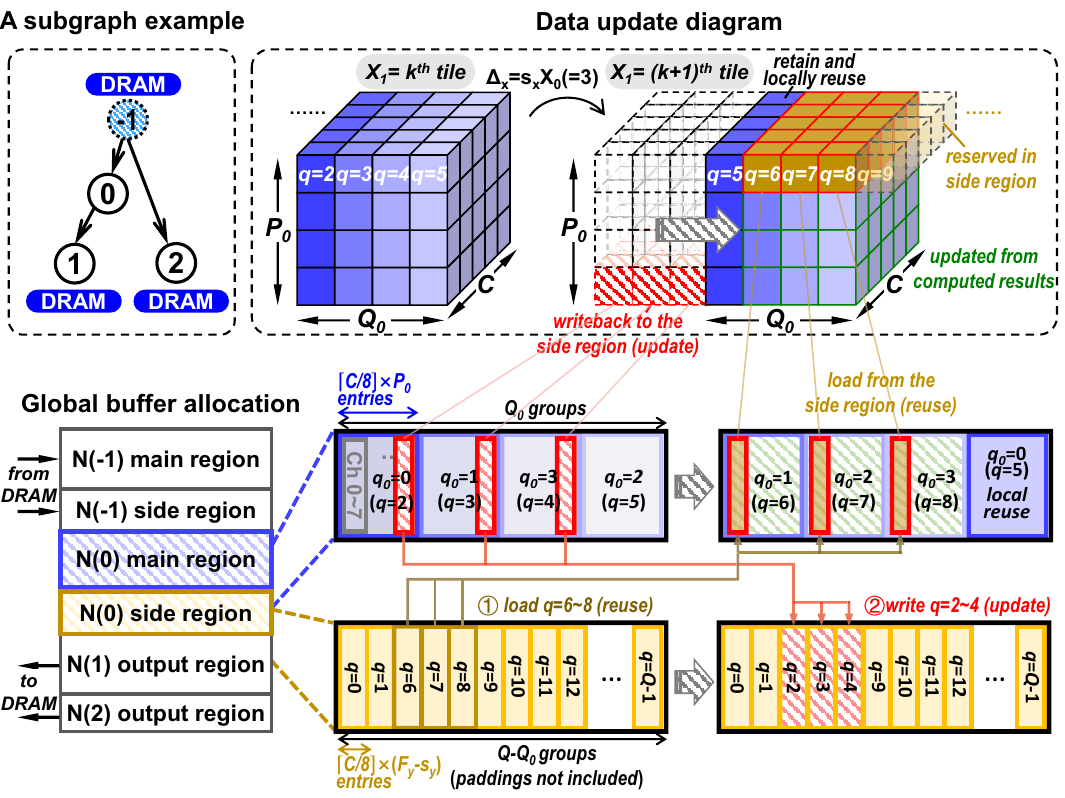}
    \caption{Memory allocation and data update scheme in the global buffer for full data reuse. The data layout used in our implementation is NWHC8c (aligned to 8 channels), which can be changed in another design. 
    $P_0$ and $Q_0$ are the height and width of an input tile; $C$ is the input channel size; $q$ is the global width-dimension index of the input tensor; and $q_0$ is the width-dimension index of an input tile.}
    \label{fig:data_layout}
    % \vspace{-2mm}
\end{figure}

Up to now, we have inferred the execution scheme for subgraphs, and the remaining challenge is how to implement it on hardware efficiently.
Figure~\ref{fig:data_layout} shows the memory allocation and update scheme for the subgraph execution.
Before computing a subgraph, the compiler determines logical blocks for input, intermediate, and output nodes, where the block sizes depend on the tile sizes derived from the execution flow.

\hl{For convenient management, we introduce two types of memory regions: \texttt{MAIN} and \texttt{SIDE}.
The \texttt{MAIN} region stores the source data for PE (i.e., the tile of $P_0\times Q_0\times C$ in Figure~\ref{fig:data_layout}).
The \texttt{SIDE} region reserves the horizontally overlapping data\footnote{We assume the column is the inner loop while the row is the outer loop.}.}
Considering no reuse requirement for some output nodes, we only need a \texttt{MAIN} region to buffer the results of the current tile.
Except for the input nodes (negative numbers) loading data from DRAM, the other nodes update data locally based on the computed results of the input node(s).

\begin{figure}[!t]
    % \vspace{-2mm}
    \centering
    \includegraphics[width=\columnwidth]{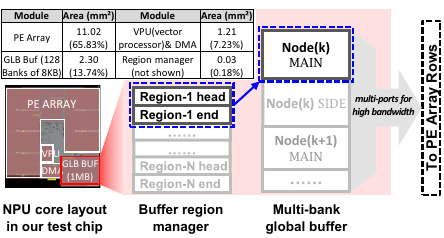}
    \caption{Hardware implementation with the buffer region manager in our 12nm NPU as a demonstration. The layout is an NPU core extracted from part of our in-house chip.}
    \label{fig:mem_arch}
    % \vspace{-2mm}
\end{figure}

In detail, the update scheme leverages the collaboration between the \texttt{MAIN} region and the \texttt{SIDE} region to achieve full reuse across sliding tiles (we consider kernel size $>$ stride).
\hl{
As shown in Figure~\ref{fig:data_layout}, when the convolution windows slide across the feature maps, the vertical overlap data (e.g., column $q=5$) are reused locally in the \texttt{MAIN} region.
In contrast, the horizontally overlapping data (e.g., the first row of $q=6\sim8$) are loaded from the \texttt{SIDE} region (path~\ding{172}).
Only a subset of data is replaced by the newly calculated results (marked in green).
Besides, the bottom horizontal slices write new data to the \texttt{SIDE} region for the next row loop (path~\ding{173}).}

The \hl{extra hardware overhead} for the proposed memory scheme is slight.
Figure~\ref{fig:mem_arch} presents our 12nm NPU core for the subgraph processing, with a buffer region manager to logically partition the global buffer to support contiguous layer processing.
The buffer region manager is a $2N$-depth register file, where $N$ determines the maximum subgraph size, and each entry pair indicates the start and the end address for each region.
The area overhead is quite small, and in our test chip, the area ratio is only 0.18\% with $N=64$ and 272-byte size (17-bit address for the 1MB 64bit-width global buffer).

\hl{In summary, our high-level idea is to divide the buffer into logical blocks for different layers and try to reuse data for sliding convolution windows. 
The memory management approach can be compatible with an accelerator as long as it supports the data movement inside the on-chip memory and flexible data assignment for computing.
Coupled with our subgraph execution scheme introduced before, intermediate outputs in the subgraph can avoid being recomputed. Only those layers required by other subgraphs are written back to DRAM for further reuse.
}
\section{Memory Communication-Capacity Co-Exploration}
\label{sec:cocco}

The aforementioned hardware model enables arbitrary subgraph execution, but there is always limited buffer capacity in hardware. 
Therefore, we need to partition the whole computation graph into a series of subgraphs that fit the memory.
Below, we move up to the optimization for graph partition and memory design-space exploration for challenge~3.

\subsection{Problem Formulation}

\subsubsection{Graph-Level Partition}
% Partitioning

Formally, a DNN model can be represented as a \textit{computation graph} $G=(V,E)$, where $V$ is the vertex set consisting of all the layers in a DNN model, and $E$ is the edge set that defines the structure of DNN.
In particular, an edge $(u,v)\in E$ represents that the output of layer $u$ is an input of layer $v$.

We aim to find a \textit{partition scheme} $P:V\rightarrow \mathbb{N}$ that assigns each layer to a subgraph, where layer $v\in V$ is computed in the $P(v)$-th subgraph.
A valid partition scheme should satisfy that any layer is computed before use.
Therefore, for any $(u,v)\in E$, we have $P(u)\leq P(v)$. Moreover, any subgraph should be connected in $G$, otherwise meaningless.

We cast the partition exploration as an optimization problem. 
The objective is to find a valid partition scheme $P$ that minimizes the total cost: 
\begin{equation}
\label{equ:partition-cost}
    \sum_i Cost_{M}(\{v\in V\mid P(v)=i\}),
\end{equation}
where $Cost_M$ is a cost function of a given subgraph based on a
target metric $M$ (e.g., external memory access (EMA) and energy).
\hl{For each subgraph, the EMA cost contains the loading of weights and input activations and the storage of output activations\footnote{The nodes that are required to write-back to DRAM can be the model output layer or the layers required by the future subgraph.}.
The energy cost includes the overhead of EMA, on-chip buffers, and computation units.}

\subsubsection{Design-Space Exploration (DSE)}
Our work further extends the optimization to combine with the memory design-space exploration.
In this paper, we focus on the global buffer and the weight buffer, given that they dominate the overhead of energy and area in an NPU core.
% \footnote{The framework also supports DSE for other hardware parameters.}
As illustrated in Figure~\ref{fig:eval}, a larger buffer capacity can take in more layers inside a subgraph, reducing communication costs but compromising the silicon area. To co-explore the hardware configuration and mapping, we construct an objective function by a linear combination of the hardware and mapping costs:
\begin{equation}
    \label{equ:cost}
    \mbox{BUF\_SIZE} + \alpha \cdot \sum_i Cost_M(\{v\in V\mid P(v)=i\}),
% \vspace{-2mm}
\end{equation}
where $\alpha$ is a preference hyper-parameter to adjust the proportion between two costs.

\subsection{Baseline Methods}
\label{subsec:baseline}

Several optimization methods that exist today can perform graph-level partition.
However, most of them fail to directly co-explore hardware and partition.
Below, we list four typical methods as our baselines and sketch their features.

\begin{figure*}[!ht]
    % \vspace{-4mm}
    \centering
    \includegraphics[width=\textwidth]{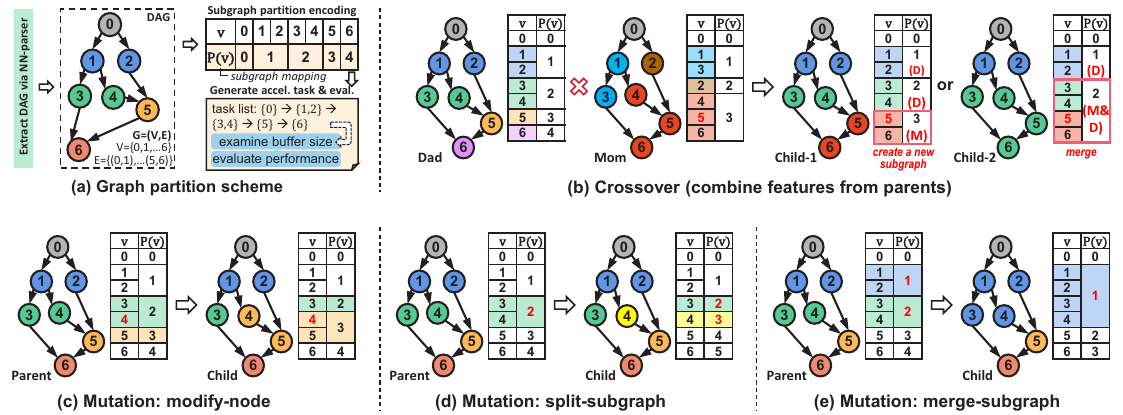}
    \caption{Illustration of crossover and mutation operations in Cocco.}
    \label{fig:genetic_operations}
    % \vspace{-1mm}
\end{figure*}

\subsubsection{Enumeration-based Algorithm}

Fused-CNN~\cite{fused-CNN} applies a straightforward way to enumerate all possible partition schemes and return the best one. Jangda \textit{et al.}~\cite{DBLP:conf/ppopp/JangdaB18} proposed state compression dynamic programming to speed up the enumeration-based algorithm.
We migrate their methods as our baseline and further improve them by only recording one subgraph in the state to reduce the time complexity.

Nonetheless, there are still exponential states in the improved implementation.
Let $N$ be the number of nodes in a graph, and the enumeration-based method may explore up to $O(2^{2N})$ states for irregular networks.
Consequently, the search is hard to complete within a reasonable search time for large-scale networks, not to mention the co-exploration with DSE.

\subsubsection{Greedy Algorithm}
Halide~\cite{Halide} employs a greedy algorithm to perform function grouping, which can be applied to the graph-level partition. 
Specifically, it first assigns each layer into a single-layer subgraph. 
Then it iteratively fuses a pair of subgraphs contributing the greatest benefit until all benefits are negative.

Therefore, this algorithm tends to be trapped at the local minimum.
Moreover, since the fusion decision rules are based on a given hardware, the greedy method cannot co-explore with DSE.

\subsubsection{Dynamic Programming (DP)-based Algorithm}
For the irregular network scheduling problem, Zheng \textit{et al.}~\cite{irregular} proposed a DP-based algorithm.
\hl{They arrange the layers based on their depth and perform DP in a sequential manner.}

\hl{This method is restricted to assigning layers that are contiguous in the depth order into a subgraph, hence the exploration is confined to constrained search space.}
It is unlikely to find the global optimum, especially for non-plain network structures. In addition, since the state transition of DP depends on the predefined buffer size, it is also tough to carry out co-exploration.

\subsubsection{Simulated Annealing (SA)}
SA~\cite{SA} is a popular optimization algorithm that samples a point and updates it iteratively to improve.
It adopts the new sample points with a probability affected by the performance difference and a hyper-parameter named temperature.
We employ the customized mutation operations (described in Section~\ref{sec:mutation}) to update the sample points and implement an SA-based algorithm as a baseline.

SA is an alternative optimization method for our framework with compatible operators, but it is not stable as the genetic algorithm in a range of benchmarks, which will be shown in later experiments.

\subsection{Genetic Algorithm}
Previous research shows competitive performance of the Genetic Algorithm (GA) in several scheduling optimization problems~\cite{GAMMA, magma}.
We summarize several benefits of GA for our hardware-mapping co-exploration problem:
\begin{enumerate}
    \item \textbf{White-box property:} We can track and tune its optimization process conveniently.
    Therefore, it is easy and intuitive to understand.
    \item \textbf{Complete search space:} It has the potential to explore the complete search space by customized mutation and crossover operations.
    \item \textbf{Avoid local optima:} In contrast to the greedy algorithm, GA can naturally jump out of the local minimum benefiting from the diversity of the population.
    \item \textbf{Flexible initialization:} We can use the results of other optimization algorithms to initialize GA and use GA to finetune the result.
    \item \textbf{Co-exploration:} Through the proposed GA operations and genome encoding, it can further support partition-DSE co-exploration.
\end{enumerate} 

We encode each candidate solution (partition scheme and the corresponding memory configuration for our problem) as a \textit{genome}, and the \textit{population} contains a set of genomes.
The GA goes through a series of \textit{generations} to obtain a lower cost.
It performs the \textit{crossover} and \textit{mutation} operations on the population in each generation.
Specifically, a crossover operation blends two genomes selected from the population to generate one offspring while a mutation operation modifies a genome randomly.
At the end of each generation, the evaluation environment evaluates the \textit{fitness} of each genome, and the population in the new generation is selected based on the fitness results.

\subsection{Cocco Optimization Framework}

Cocco is a GA-based optimization framework that enables the co-exploration of memory configuration and graph-level partition, as shown in Figure~\ref{fig:framework}.
The core of Cocco is a series of operations that explore a complete search space.
We build a genetic algorithm based on these customized operations.
Fed with the neural network structure and DSE requirements, Cocco goes through several steps to get the optimization results.
The execution model described in Section~\ref{sec:memory} is embedded in the evaluation environment.
In the following, we introduce the five stages of Cocco.

\subsubsection{Initialization}
The first step in Cocco is to generate the initial population, where each genome contains a partition scheme of the computation graph and a memory configuration for DSE.
For the DSE part, every genome selects a capacity value in a given range following a uniform distribution.
There are two options in Cocco to initialize the partition scheme $P$ of each genome.
The first option is random initialization.
Precisely, we determine the $P(v)$ for each layer $v\in V$ in topological order, and each $P(v)$ is selected randomly within the valid range.
The other option is to initialize the partition scheme from other optimization algorithms.

\subsubsection{Crossover}

We designed a customized crossover operation to inherit and blend the features of two parents selected from the population.
Specifically, each hardware configuration (i.e., memory capacity) in the offspring is the average of its parents and then rounds to the nearest candidate value.
For the partition scheme, we assign layers to subgraphs in topological order.
Each undecided layer chooses one parent randomly to reproduce the corresponding subgraph.
If the reproduced subgraph contains layers that have been decided, we split out a new one excluding those layers, or merge it with one of the subgraphs to which the decided layers belong.
% It depends on a hyper-parameter \textit{merge\_rate}.

As shown in Figure~\ref{fig:genetic_operations}(b), layer $1$ and layer $3$ select Dad as the parent to reproduce the subgraphs $\{1,2\}$ and $\{3,4\}$, respectively.
Next, layer $5$ selects Mom as its parent, so it intends to reproduce subgraph $\{4,5,6\}$. However, since we have already decided on layer $4$ in subgraph $\{3,4\}$, there are two alternatives: creating a new subgraph $\{5,6\}$ (Child-1) or merging with subgraph $\{3,4\}$ to obtain $\{3,4,5,6\}$ (Child-2).

\begin{figure}[!t]
    % \vspace{-2mm}
    \centering
    \includegraphics[width=\columnwidth]{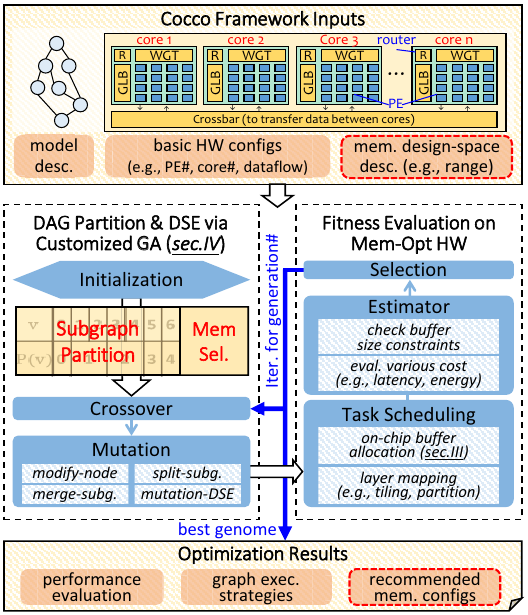}
    \caption{Cocco framework overview.}
    \label{fig:framework}
    % \vspace{-4mm}
\end{figure}

\subsubsection{Mutation}
\label{sec:mutation}

Four mutation operations are customized for the optimization flow to explore the search space extensively.
We guarantee the validity of genomes after each mutation in the implementation.
At the bottom of Figure~\ref{fig:genetic_operations}, we show a node-level operation (\texttt{modify-node}) and two subgraph-level ones (\texttt{split-subgraph} and \texttt{merge-subgraph}):
% \footnote{All these operations guarantee the validity of genomes after mutation.}
\begin{itemize}
    \item \texttt{modify-node}~(Figure~\ref{fig:genetic_operations}(c)): Modify the assignment of a randomly selected node $u$: from $u\to P(u)$ to $u\to P'(u)$, where $P'(u)$ can be an existed subgraph or a new one.
    \item \texttt{split-subgraph}~(Figure~\ref{fig:genetic_operations}(d)): Split a randomly selected subgraph into two or more subgraphs.
    \item \texttt{merge-subgraph}~(Figure~\ref{fig:genetic_operations}(e)): Merge two randomly selected subgraphs into one subgraph.
    \item \texttt{mutation-DSE}~(not shown): Modify the memory configuration to a random one within the range.
    The new values are sampled based on a normal distribution, where the average is the original value, and the variance is a hyper-parameter.
\end{itemize}

\subsubsection{Evaluation}

Since GA tries to maximize the fitness of the genomes, we set fitness to be the opposite of the cost (e.g., Formula~\ref{equ:partition-cost} and~\ref{equ:cost}).
To evaluate the fitness of each genome in the population, we use our simulator (introduced in the next section) to extract the execution costs of subgraphs (e.g., EMA and energy).

During the evaluation, the simulator decodes the subgraph and hardware configuration of each genome and calculates the fitness by aggregating the cost of each subgraph.
Particularly, when a large subgraph exceeds the buffer capacity, we perform the \texttt{split-subgraph} operation to ensure genome validity.
This kind of in-situ tuning can increase the number of valid samples during the optimization operations and thus, improve the sample efficiency.

\subsubsection{Selection}
At the end of each generation, Cocco performs the tournament selection.
Specifically, it holds multiple tournaments among a few randomly selected genomes, and the winners (the genome with the best fitness) of these tournaments form the population of a new generation.
This operation facilitates superior fitness in the new generation.
The number of genomes in each tournament is decided by a hyper-parameter.
% Specifically, the genomes with the best fitness will form an elite set and stay unchanged in the new generation.
The new generation subsequently starts from the crossover step again.
\section{Experiments}
\label{sec:eval}

In the evaluations, we first present the superiority of Cocco for the graph partition; and then demonstrate its outstanding stability and sample efficiency of the co-exploration for the hardware optimization, followed by additional discussions about the results under different configurations.

\subsection{Methodology}

\subsubsection{Evaluated Models}

In the following evaluations, we consider three types of model structures: plain (VGG16~\cite{vgg}), multi-branch (ResNet50, ResNet152~\cite{res}, GoogleNet~\cite{inception}, Transformer~\cite{transformer}, and GPT~\cite{GPT}), and irregular structure (RandWire-A/B~\cite{random-gen} and NasNet~\cite{nasnet}).
RandWire-A/B are generated based on the small and regular regime configurations introduced in the paper~\cite{random-gen}.
% We choose these models for studying various complexity of network structures
FC layers are transformed to 1$\times$1 CONV while pooling and element-wise layers are analyzed as depth-wise CONV without weights.
The scalar operations (e.g., activation function) are hidden in the pipeline (e.g., the post-process module following PE in Simba~\cite{simba}) and their overhead can be ignored.

\begin{figure*}[!t]
    % \vspace{-2mm}
    \centering
    \includegraphics[width=\textwidth]{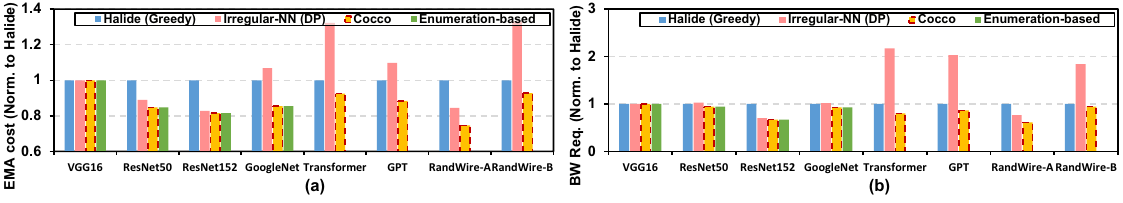}
    \caption{The evaluation results for graph partition using the EMA-opt configuration (EMA as the optimization metric).
    The enumeration-based method is deterministic, which figures out the optimal solution as a reference in the first four models.
    It cannot complete for large-scale models (Transformer, GPT, RandWire-A, and RandWire-B) in a reasonable time because of the exponential search space.}
    \label{fig:exp1}
    % \vspace{-1mm}
\end{figure*}

\subsubsection{Accelerator Platform}

As shown at the top of Figure~\ref{fig:framework}, we \hl{consider a SIMBA-like hierarchical accelerator with} a global buffer, a weight buffer, and a 4$\times$4 PE array in each core used in several previous works~\cite{nnbaton, simba, amos}.
Each PE contains an 8$\times$8 MAC array to process a sub-tile from the global buffer.
In particular, we model the execution flow based on the scheme described in Section~\ref{sec:memory}.
% the memory execution flow is modeled with the features described in Section~\ref{sec:memory}.
The parallelism of two dimensions of the PE array can be dynamically configured by the mapper results to ensure high utilization.
We schedule subgraphs in topological order and prefetch weights of the next subgraph during the current computing.
We also extend our platform to support fundamental multi-core studies by interconnecting cores with a crossbar.
They share weights to release the burden of each core.

The arithmetic and memory overhead is extracted in a 12nm library based on the synthesized RTL implementations (SRAM based on the ARM memory compiler) with 1GHz.
The DRAM energy is set as 12.5pJ/bit~\cite{interstellar}.
The extra footprint of the plug-in design is mainly a 272-Byte register file to store the head and end logical region addresses of maximal 64 nodes, which is negligible.
Based on off-the-shelf evaluators Timeloop~\cite{timeloop} and MAESTRO~\cite{understand} for spatial accelerators, we developed a modified simulator that supports the evaluation of latency and energy.
It employs the consumption-centric scheme to determine the tile size of each layer, and the memory access in the model is free from padding data.
The latency per subgraph depends on the maximum of the calculation and external communication cycles.
\hl{We allocate 16GB/s DRAM bandwidth per accelerator core for loading weights and input activations and writing back data for subsequent subgraphs.
The off-chip communication consists of weight loading of each layer and the inputs and outputs of each subgraph.
As described in Section~\ref{sec:memory}, our subgraph execution scheme avoids recomputing of intermediate outputs.
}

\subsubsection{Baselines}
Three optimization baselines for graph partition are the greedy algorithm used in Halide~\cite{Halide}, dynamic programming (DP) used in Irregular-NN~\cite{irregular}
% ~\footnote{Considering that the implementation of its customized DP lacks details, we arrange layers based on the depth as they described.}
, and the enumeration-based method as a reference.

For the DSE studies, we compare Cocco with simulated annealing (SA)~\cite{SA} to demonstrate the better stability of GA.
These two methods are both the co-optimization scheme that optimizes partition and hardware settings at the same time.
In contrast to co-optimization, the two-step scheme is another method for design-space exploration.
Specifically, we use random search (RS) or grid search (GS) to sample memory capacity candidates and then explore the corresponding partition schemes.
During the search, we evaluate 5,000 samples for each capacity candidate and keep the best candidate as the output.
As for the sampling method, RS randomly samples memory capacity candidates while GS uses a coarser granularity to enumerate the candidates.

\subsection{Graph Partition Evaluations}

We start by presenting the partition performance on the single-core hardware with a 1MB global buffer and a 1.125MB weight buffer.
The number of samples in Cocco is set to be 400,000.
We evaluate the external memory access (EMA) and bandwidth requirements of eight models shown in Figure~\ref{fig:exp1}, where the results are normalized to the Halide baseline.
This experiment aims to validate the effectiveness of our Cocco framework in graph partition.
For networks with simpler structures, Cocco can find out the optimal solutions same as the enumeration-based results.
For large-scale irregular networks (Transformer, GPT, RandWire-A, and RandWire-B), the enumeration-based method cannot complete in a reasonable time, while Cocco provides better solutions than Halide and DP.
A better subgraph partition strategy helps to ease the communication burden, thus reducing the EMA cost and bandwidth requirements.
% VGG16 is a small plain network so that all optimization methods can easily find the optimum.
% When the structure becomes complex, greedy search and DP start to fall behind due to the local minimum problem and constraint search space, respectively.

\begin{table}[!t]
\centering
\caption{Hardware-mapping co-exploration for separate buffer. In this table, A refers to the global buffer, and W refers to the weight buffer. We evaluate the cost using Formula~\ref{equ:cost} (the lower cost, the better), where the metric $M$ is energy. We use RandWire-A as RandWire in the following experiments.}
\label{tab:exp2-2-separate}
\resizebox{\columnwidth}{!}{%
\setlength{\tabcolsep}{1.3mm}{
\begin{tabular}{@{}cc|ccc|ccc@{}}
\toprule
\multicolumn{2}{c|}{\multirow{2.5}{*}{\textbf{Optimization }}} &
  \multicolumn{3}{c|}{\textbf{ResNet50}} &
  \multicolumn{3}{c}{\textbf{GoogleNet}}
  \\ \cmidrule(l){3-8} 
\multicolumn{2}{c|}{} &
  \textbf{Size (A)} &
  \textbf{Size (W)} &
  \textbf{Cost} &
  \textbf{Size (A)} &
  \textbf{Size (W)} &
  \textbf{Cost}
  \\ \midrule
\multicolumn{1}{c|}{\multirow{4}{*}{\textbf{\begin{tabular}[c]{@{}c@{}}Fixed\\HW
\end{tabular}}}} &
  \textbf{Buf(S)}  & 512KB
   & 576KB
   & \textbf{1.04E7} %10308821
   & 512KB
   & 576KB
   & 4.07E6 %4074382
   \\ \cmidrule(r){2-8}
\multicolumn{1}{c|}{} &
  \textbf{Buf(M)} & 1024KB
   & 1152KB
   & 1.07E7 %10700399
   & 1024KB
   & 1152KB
   & 5.06E6 %5062113
   \\ \cmidrule(r){2-8}
\multicolumn{1}{c|}{} &
  \textbf{Buf(L)} & 2048KB
   & 2304KB
   & 1.24E7 %12426863
   & 2048KB
   & 2304KB
   & 7.18E6 %7178382
   \\ \midrule
\multicolumn{1}{c|}{\multirow{2.6}{*}{\textbf{\begin{tabular}[c]{@{}c@{}}Two-Step
\end{tabular}}}} &
  \textbf{RS+GA} & 448KB
   & 864KB
   & \textbf{1.04E7} %10437845
   & 384KB
   & 432KB
   & 3.88E6 % 3884603
   \\ \cmidrule(r){2-8}
\multicolumn{1}{c|}{} &
  \textbf{GS+GA} & 128KB
   & 864KB
   & 1.07E7 %10692207
   & 128KB
   & 144KB
   & 3.80E6 %3799201
   \\ \midrule
\multicolumn{1}{c|}{\multirow{2.6}{*}{\textbf{\begin{tabular}[c]{@{}c@{}} Co-Opt\end{tabular}}}} &
  \textbf{SA} & 256KB
   & 360KB
   & 1.06E7 %10573423
   & 192KB
   & 144KB
   & 3.78E6 %3784456
   \\ \cmidrule(r){2-8}
\multicolumn{1}{c|}{} &
  \textbf{Cocco} & 704KB
   & 864KB
   & \cellcolor[HTML]{EFEFEF}\textbf{1.04E7} %10378863
   & 192KB
   & 432KB
   & \cellcolor[HTML]{EFEFEF}\textbf{3.75E6} %3748206
   \\
   \toprule
   \toprule
\multicolumn{2}{c|}{\multirow{2.5}{*}{\textbf{Optimization}}} &
  \multicolumn{3}{c|}{\textbf{RandWire}} &
  \multicolumn{3}{c}{\textbf{NasNet}}
  \\ \cmidrule(l){3-8} 
\multicolumn{2}{c|}{} &
  \textbf{Size (A)} &
  \textbf{Size (W)} &
  \textbf{Cost} &
  \textbf{Size (A)} &
  \textbf{Size (W)} &
  \textbf{Cost}
  \\ \midrule
\multicolumn{1}{c|}{\multirow{4}{*}{\textbf{\begin{tabular}[c]{@{}c@{}}Fixed\\HW
\end{tabular}}}} &
  \textbf{Buf(S)}  & 512KB
   & 576KB
   & 3.23E6 %3236246
   & 512KB
   & 576KB
   & 6.14E7 %61408649
   \\ \cmidrule(r){2-8}
\multicolumn{1}{c|}{} &
  \textbf{Buf(M)} & 1024KB
   & 1152KB
   & 3.92E6 %3916179
   & 1024KB
   & 1152KB
   & 5.83E7 %58315482
   \\ \cmidrule(r){2-8}
\multicolumn{1}{c|}{} &
  \textbf{Buf(L)} & 2048KB
   & 2304KB
   & 6.00E6 %6003241
   & 2048KB
   & 2304KB
   & 5.66E7 %56600470
   \\ \midrule
\multicolumn{1}{c|}{\multirow{2.6}{*}{\textbf{\begin{tabular}[c]{@{}c@{}}Two-Step
\end{tabular}}}} &
  \textbf{RS+GA} & 448KB
   & 792KB
   & 3.31E6 %3309339
   & 1152KB
   & 2016KB
   & 5.60E7 %55960808
   \\ \cmidrule(r){2-8}
\multicolumn{1}{c|}{} &
  \textbf{GS+GA} & 128KB
   & 144KB
   & 3.02E6 %3021982
   & 2048KB
   & 2304KB
   & 5.66E7 %56600470
   \\ \midrule
\multicolumn{1}{c|}{\multirow{2.6}{*}{\textbf{\begin{tabular}[c]{@{}c@{}} Co-Opt\end{tabular}}}} &
  \textbf{SA} & 192KB
   & 144KB
   & 3.00E6 %2998848
   & 2048KB
   & 1872KB
   & 5.61E7 %56060305
   \\ \cmidrule(r){2-8}
\multicolumn{1}{c|}{} &
  \textbf{Cocco} & 256KB
   & 144KB
   & \cellcolor[HTML]{EFEFEF}\textbf{2.98E6} %2981828
   & 1280KB
   & 2088KB
   & \cellcolor[HTML]{EFEFEF}\textbf{5.59E7} %55874290
   \\
   \bottomrule
\end{tabular}%
}}
\vspace{-3mm}
\end{table}

\subsection{Hardware-Mapping Co-Exploration}
After learning the superiority of Cocco for the graph partition, we further co-explore the memory configuration and graph partition mapping as the core study of this work.
Three categories of exploration methods are used, including the \textit{fixed hardware scheme}, the \textit{two-step scheme} as baselines, and the proposed \textit{co-optimization scheme}.
We set three fixed memory configurations with \underline{S}mall capacity, \underline{M}edium capacity, and \underline{L}arge capacity, followed by a partition-only procedure.
The two-step scheme is implemented with decoupled steps for capacity search (RS or GS) and partition (GA).
The co-optimization methods include the proposed Cocco and an SA-based one as the comparison.
All methods sample up to 50,000 points.
The energy-capacity co-optimization is used in the following evaluations.

\begin{table}[!t]
% \vspace{-2mm}
\centering
\caption{Hardware-mapping co-exploration for shared buffer. We evaluate the cost using Formula~\ref{equ:cost} (the lower cost, the better), where the metric $M$ is energy.}
\label{tab:exp2-2-shared}
\resizebox{\columnwidth}{!}{%
\setlength{\tabcolsep}{4.1mm}{
\begin{tabular}{@{}cc|cc|cc@{}}
\toprule
\multicolumn{2}{c|}{\multirow{2.5}{*}{\textbf{Optimization}}} &
  \multicolumn{2}{c|}{\textbf{ResNet50}} &
  \multicolumn{2}{c}{\textbf{GoogleNet}}
  \\ \cmidrule(l){3-6} 
\multicolumn{2}{c|}{} &
  \textbf{Size} &
  \textbf{Cost} &
  \textbf{Size} &
  \textbf{Cost} 
  \\ \midrule
\multicolumn{1}{c|}{\multirow{4}{*}{\textbf{\begin{tabular}[c]{@{}c@{}}Fixed\\HW
\end{tabular}}}} &
  \textbf{Buf(S)}  & 576KB
   & 1.01E7 %10080455
   & 576KB
   & 3.66E6 %3658762
   \\ \cmidrule(r){2-6}
\multicolumn{1}{c|}{} &
  \textbf{Buf(M)} & 1152KB
   & 1.00E7 %10002365
   & 1152KB
   & 4.04E6 %4039880
   \\ \cmidrule(r){2-6}
\multicolumn{1}{c|}{} &
  \textbf{Buf(L)} & 2304KB
   & 1.04E7 %10359126
   & 2304KB
   & 5.09E6 %5090638
   \\ \midrule
\multicolumn{1}{c|}{\multirow{2.6}{*}{\textbf{\begin{tabular}[c]{@{}c@{}}Two-Step
\end{tabular}}}} &
  \textbf{RS+GA} & 1280KB
   & \textbf{0.98E7} %9812310
   & 640KB
   & 3.65E6 %3652321
   \\ \cmidrule(r){2-6}
\multicolumn{1}{c|}{} &
  \textbf{GS+GA} & 1344KB
   & \textbf{0.98E7} %9797565
   & 512KB
   & 3.65E6 %3654343
   \\ \midrule
\multicolumn{1}{c|}{\multirow{2.6}{*}{\textbf{\begin{tabular}[c]{@{}c@{}} Co-Opt\end{tabular}}}} &
  \textbf{SA} & 896KB
   & 1.00E7 %9981065
   & 192KB
   & 3.75E6 %3752554
   \\ \cmidrule(r){2-6}
\multicolumn{1}{c|}{} &
  \textbf{Cocco} & 1344KB
   & \cellcolor[HTML]{EFEFEF}\textbf{0.98E7} %9797565
   & 384KB
   & \cellcolor[HTML]{EFEFEF}\textbf{3.60E6} %3597837
   \\ 
   \toprule
   \toprule
\multicolumn{2}{c|}{\multirow{2.5}{*}{\textbf{Optimization}}} &
  \multicolumn{2}{c|}{\textbf{RandWire}} &
  \multicolumn{2}{c}{\textbf{NasNet}}
  \\
   \cmidrule(l){3-6} 
\multicolumn{2}{c|}{} &
  \textbf{Size} &
  \textbf{Cost} &
  \textbf{Size} &
  \textbf{Cost} \\ \midrule
\multicolumn{1}{c|}{\multirow{4}{*}{\textbf{\begin{tabular}[c]{@{}c@{}}Fixed\\HW
\end{tabular}}}} &
  \textbf{Buf(S)}
   & 576KB
   & \textbf{2.83E6} %2826426
   & 576KB
   & 6.36E7 %63644119
   \\ \cmidrule(r){2-6}
\multicolumn{1}{c|}{} &
  \textbf{Buf(M)}
   & 1152KB
   & 3.03E6 %3030616
   & 1152KB
   & 5.73E7 %57311782
   \\ \cmidrule(r){2-6}
\multicolumn{1}{c|}{} &
  \textbf{Buf(L)}
   & 2304KB
   & 3.90E6 %3901447
   & 2304KB
   & 5.51E7 %55108316
   \\ \midrule
\multicolumn{1}{c|}{\multirow{2.6}{*}{\textbf{\begin{tabular}[c]{@{}c@{}}Two-Step
\end{tabular}}}} &
  \textbf{RS+GA}
   & 320KB
   & 2.85E6 %2847884
   & 2560KB
   & 5.47E7 %54743556
   \\ \cmidrule(r){2-6}
\multicolumn{1}{c|}{} &
  \textbf{GS+GA}
   & 832KB
   & 2.86E6 %2864989
   & 3072KB
   & 5.42E7 %54169038
   \\ \midrule
\multicolumn{1}{c|}{\multirow{2.6}{*}{\textbf{\begin{tabular}[c]{@{}c@{}} Co-Opt\end{tabular}}}} &
  \textbf{SA}
   & 256KB
   & 2.92E6 %2922746
   & 1728KB
   & 5.56E7 %55552213
   \\ \cmidrule(r){2-6}
\multicolumn{1}{c|}{} &
  \textbf{Cocco}
   & 384KB
   & \cellcolor[HTML]{EFEFEF}\textbf{2.83E6} %2831818
   & 2624KB
   & \cellcolor[HTML]{EFEFEF}\textbf{5.37E7} %53689816
   \\
   
   \bottomrule
   
\end{tabular}%
}}
% \vspace{-6mm}
\end{table}

\subsubsection{DSE analysis using separate and shared buffer}
We first perform the hardware-mapping co-exploration to determine the {suitable} memory configuration (except for the fixed-HW scheme) with $\alpha=0.002$\footnote{The energy and the capacity units are pJ and Byte, respectively.} and then solely execute the partition-only Cocco to obtain the final cost. %(RandWire-A is used in this experiment)
In particular, we also compared the results using two memory designs: separate buffer and shared buffer.
For the separate buffer design, activations and weights are stored in different buffers while they share the same space in the shared buffer design.
The memory capacity candidates for the global buffer (for activations) range from 128KB to 2048KB with a 64KB interval, while that for the weight buffer range from 144KB to 2304KB with a 72KB interval.
The exploration range of the shared buffer is from 128KB to 3072KB with an interval of 64KB.

The evaluation using separate buffers is shown in Table~\ref{tab:exp2-2-separate}, where Cocco achieves better optimization with up to 1.89\% (compared to SA in ResNet50) to 50.33\% (compared to Fixed-HW(L) in RandWire) lower cost compared to various baselines across four models.
% The fixed-HW methods sometimes achieve good performance when the capacity is close to the optimum.
The two-step scheme fails to combine the information between different sizes, so it is generally worse than the co-optimization method.

The capacity results also reflect the inherent capacity preference of different models.
The data amount in GoogleNet and RandWire is relatively smaller, and thus their capacity requirements are lower.
In contrast, the data amount in NasNet is larger, so a high capacity is preferred.

As shown in Table~\ref{tab:exp2-2-shared}, the evaluation of the shared buffer setting shows a similar trend.
Furthermore, we can observe that most of the cost results of the shared buffer are lower than those using the separate configuration.
Although the shared buffer design requires additional control flows, it indeed improves efficiency than the separate buffer design.
% For ease of implementation, the NPU architect tends to employ a separate-buffer design for activation and weight, but the shared buffer design indeed improves efficiency.

\begin{figure}[!t]
    % \vspace{-1mm}
    \centering
    \includegraphics[width=\columnwidth]{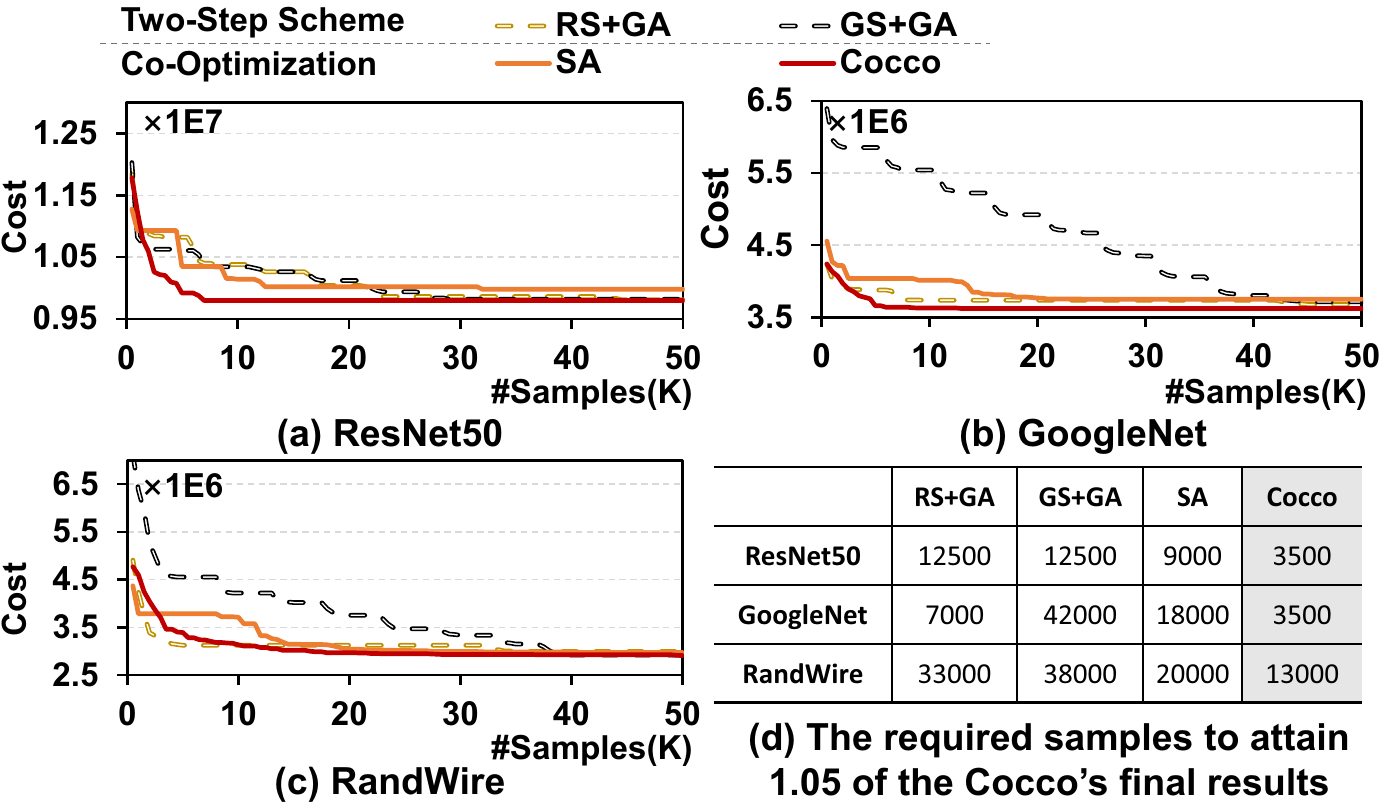}
    \caption{The convergence curve of Cocco compared with other baselines in the hardware-mapping co-explorations.
    The optimization method requiring fewer samples in (d) has higher sample efficiency.}
    \label{fig:exp2-1}
    % \vspace{-2mm}
\end{figure}
\begin{figure}[!t]
    % \vspace{-4mm}
    \centering
    \includegraphics[width=\columnwidth]{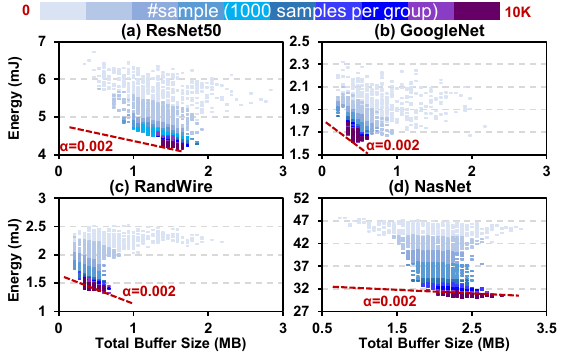}
    \caption{The visualization of sample points distribution during optimization. The slope of the red dashed line denotes the preference between energy and capacity cost. The point on the line with a lower intercept has a smaller cost.}
    \label{fig:exp2-3}
    % \vspace{-4mm}
\end{figure}

\subsubsection{Sample efficiency analysis}

We next study the sample efficiency of the two-step and the co-optimization scheme in Figure~\ref{fig:exp2-1}.
We record the cost trends of the first 50,000 samples on ResNet50, GoogleNet, and RandWire during the exploration.
% (the subsequent partition-only stage for fine-tuning is not included).
Overall, Cocco shows a consistent convergence trend on these three networks.
And it converges faster and achieves lower costs compared to other baselines, exhibiting a higher sample efficiency.
The two-step methods perform graph-partition separately under different capacities, so they fail to utilize the partition information between capacities.
Particularly, the GS method uses a deterministic search direction (search from large to small capacity in this experiment), so the convergence time depends on the optimal capacity.
Since GoogleNet and RandWire require relatively small buffers, GS takes a considerable number of samples to converge.

\subsubsection{Optimization procedure analysis}

We next study how the distribution of sample points changes during the optimization procedure of Cocco.
While searching for 20 generations with 500 genomes each, we divided them into ten groups with different colors in Figure~\ref{fig:exp2-3}.
The results show that the distribution moves towards a lower intercept of the $\alpha$-slope line and gets more centralized in the later generations during the optimization process of Cocco.
\begin{figure}[!t]
    % \vspace{-4mm}
    \centering
    \includegraphics[width=\columnwidth]{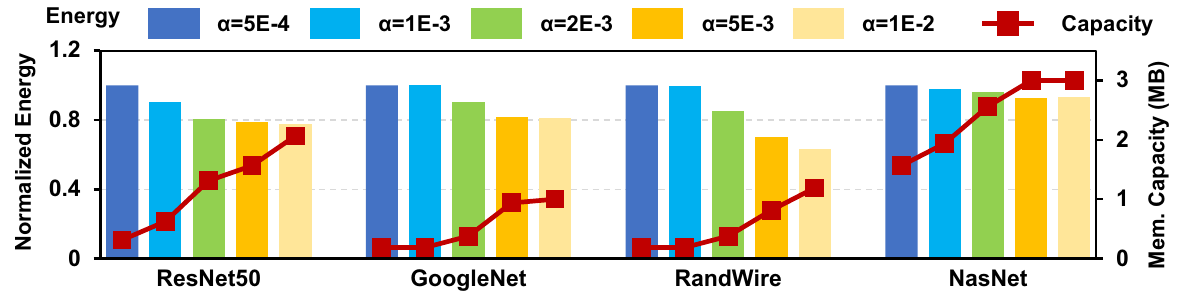}
    \caption{The trade-off between energy and memory capacity.
    The optimization target is to minimize the cost defined in Formula~\ref{equ:cost}, where the metric $M$ is energy.
    % (using the energy-capacity co-opt configuration).
    Energy results of each model are normalized to the first $\alpha(=0.0005)$ results.}
    \label{fig:exp3-1-1}
    % \vspace{-2mm}
\end{figure}

\subsection{Sensitivity Study about Cocco framework}

\subsubsection{Study of $\alpha$ in the cost function}

% In the cost function, we introduce a preference hyper-parameter $\alpha$ to adjust the proportion of two costs.
The results shown in Figure~\ref{fig:exp3-1-1} demonstrate the effectiveness of $\alpha$ in adjusting the preference between the memory capacity and the given metric (energy is used here).
The optimization trades the memory capacity for lower energy cost with the increase of $\alpha$.
In addition, a larger memory capacity indeed contributes to lower energy, but the yields show differences because of their various model-inherent graph and layer patterns.
For example, NasNet is more memory-intensive and more structure-complex than the other three models, so it requires a larger memory capacity for less energy consumption.

\begin{table}[!t]
\centering
\caption{Multi-core and batch evaluation using the energy-capacity co-opt configuration.
Size denotes the shared buffer size in each core.}
\label{tab:exp3-2}
\resizebox{\columnwidth}{!}{%
{
\begin{tabular}{@{}c|c|ccc|ccc@{}}
\toprule
\multicolumn{1}{c|}
{\multirow{2.5}{*}{\textbf{Core\#}}} &
\multicolumn{1}{c|}
{\multirow{2.5}{*}{\textbf{Batch}}} &
  \multicolumn{3}{c|}{\textbf{ResNet50}} &
  \multicolumn{3}{c}{\textbf{GoogleNet}}
  \\ \cmidrule(l){3-8} 
\multicolumn{1}{c|}{} &
\multicolumn{1}{c|}{} &
  \textbf{Energy(mJ)} &
  \textbf{Lat.(ms)} &
  \textbf{Size(KB)} &
  \textbf{Energy(mJ)} &
  \textbf{Lat.(ms)} &
  \textbf{Size(KB)}
  \\ \midrule
\multicolumn{1}{c|}{\multirow{4}{*}{\textbf{\begin{tabular}[c]{@{}c@{}}1
\end{tabular}}}} &
  \textbf{1}
   & 4.21
   & 4.59
   & 1344
   & 1.61
   & 2.05
   & 384
%   &
   \\ \cmidrule(r){2-8}
\multicolumn{1}{c|}{} &
  \textbf{2}
   & 6.32
   & 8.98
   & 1728
   & 2.18
   & 3.91
   & 896
%   &
   \\ \cmidrule(r){2-8}
\multicolumn{1}{c|}{} &
  \textbf{8}
   & 11.88
   & 35.93
   & 2880
   & 5.64
   & 15.53
   & 1472
%   &
   \\ \midrule
\multicolumn{1}{c|}{\multirow{4}{*}{\textbf{\begin{tabular}[c]{@{}c@{}}2
\end{tabular}}}} &
  \textbf{1}
   & 4.38
   & 2.48
   & 768
   & 1.66
   & 1.04
   & 192
%   &
   \\ \cmidrule(r){2-8}
\multicolumn{1}{c|}{} &
  \textbf{2}
   & 6.46
   & 4.78
   & 1088
   & 2.34
   & 1.99
   & 384
%   &
   \\ \cmidrule(r){2-8}
\multicolumn{1}{c|}{} &
  \textbf{8}
   & 13.01
   & 19.12
   & 1664
   & 5.84
   & 7.97
   & 960
%   &
   \\ \midrule
\multicolumn{1}{c|}{\multirow{4}{*}{\textbf{\begin{tabular}[c]{@{}c@{}} 4\end{tabular}}}} &
  \textbf{1}
   & 4.29
   & 1.39
   & 448
   & 1.34
   & 0.54
   & 192
%   &
   \\ \cmidrule(r){2-8}
\multicolumn{1}{c|}{} &
  \textbf{2}
   & 6.58
   & 2.68
   & 640
   & 2.20
   & 1.07
   & 192
%   &
   \\ \cmidrule(r){2-8}
\multicolumn{1}{c|}{} &
  \textbf{8}
   & 11.50
   & 10.71
   & 1664
   & 6.24
   & 4.30
   & 448
%   &
   \\ 
   \toprule
   \toprule
\multicolumn{1}{c|}
{\multirow{2.5}{*}{\textbf{Core\#}}} &
\multicolumn{1}{c|}
{\multirow{2.5}{*}{\textbf{Batch}}} &
  \multicolumn{3}{c|}{\textbf{RandWire}} &
  \multicolumn{3}{c}{\textbf{NasNet}}
  \\ \cmidrule(l){3-8} 
\multicolumn{1}{c|}{} &
\multicolumn{1}{c|}{} &
  \textbf{Energy(mJ)} &
  \textbf{Lat.(ms)} &
  \textbf{Size(KB)} &
  \textbf{Energy(mJ)} &
  \textbf{Lat.(ms)} &
  \textbf{Size(KB)}
  \\ \midrule
\multicolumn{1}{c|}{\multirow{4}{*}{\textbf{\begin{tabular}[c]{@{}c@{}}1
\end{tabular}}}} &
  \textbf{1}
   & 1.26
   & 1.47
   & 384
   & 28.57
   & 49.92
   & 2624
%   &
   \\ \cmidrule(r){2-8}
\multicolumn{1}{c|}{} &
  \textbf{2}
   & 2.25
   & 2.74
   & 704
   & 47.68
   & 99.87
   & 3072
%   &
   \\ \cmidrule(r){2-8}
\multicolumn{1}{c|}{} &
  \textbf{8}
   & 8.66
   & 10.85
   & 1664
   & 133.03
   & 396.90
   & 3072
%   &
   \\ \midrule
\multicolumn{1}{c|}{\multirow{4}{*}{\textbf{\begin{tabular}[c]{@{}c@{}}2
\end{tabular}}}} &
  \textbf{1}
   & 1.41
   & 0.95
   & 192
   & 29.18
   & 24.93
   & 1728
%   &
   \\ \cmidrule(r){2-8}
\multicolumn{1}{c|}{} &
  \textbf{2}
   & 2.37
   & 1.80
   & 384
   & 48.80
   & 49.73
   & 2624
%   &
   \\ \cmidrule(r){2-8}
\multicolumn{1}{c|}{} &
  \textbf{8}
   & 8.39
   & 7.16
   & 1280
   & 153.25
   & 227.19
   & 3072
%   &
   \\ \midrule
\multicolumn{1}{c|}{\multirow{4}{*}{\textbf{\begin{tabular}[c]{@{}c@{}} 4\end{tabular}}}} &
  \textbf{1}
   & 1.39
   & 0.71
   & 192
   & 28.00
   & 14.56
   & 960
%   &
   \\ \cmidrule(r){2-8}
\multicolumn{1}{c|}{} &
  \textbf{2}
   & 2.91
   & 1.40
   & 192
   & 45.03
   & 28.58
   & 1664
%   &
   \\ \cmidrule(r){2-8}
\multicolumn{1}{c|}{} &
  \textbf{8}
   & 9.24
   & 5.55
   & 960
   & 131.98
   & 133.38
   & 2816
   \\ 
   \bottomrule
\end{tabular}%
}}
% \vspace{-4mm}
\end{table}

\subsubsection{Study of performance v.s. memory capacity}

% Although this paper focuses on the single-core design, our framework can serve as a primitive tool for a multi-core system.
Figure~\ref{fig:survey} shows that the increase of capacity is sub-linear with performance.
To study this observation, we scale our model to the multi-core version and share weights of a subgraph across cores.
Different cores only buffer a subset of weights and transfer the data between cores, similar to BSD in Tangram~\cite{tangram} or data-rotation in NN-Baton~\cite{nnbaton}.
The overhead of the interconnection crossbar is extracted from the implemented Arteries IP~\cite{arteries}.

An accelerator with more cores can cover a larger subgraph but bring more core-to-core overhead.
As shown in Table~\ref{tab:exp3-2}, in most cases, energy increases from the single-core to dual-core configuration because of the communication overhead.
Moreover, profiting from the data-sharing mechanism, the required memory of each core drops with the increase of core number.

\subsubsection{Batch size study}

For the batch size evaluation shown in Table~\ref{tab:exp3-2}, the latency with a larger batch size principally presents a sub-linear increase, which benefits from the lower bandwidth requirement of weights via the inter-sample data reuse.
In addition, such data reuse amortizes the energy burden per batch processing.
And owing to the better weight reuse in multi-batch processing, a larger batch size does not require a proportional capacity.
\section{Related Works}

% \subsection{Memory Optimization for DNN accelerators}
% As a specialized processor for tensor computing, DNN accelerators need to manage a chunk of data via reasonable space allocation for workload assignment and manage the access/update sequence for data reuse.

\subsection{Intra-layer Optimization}

Prior works focus on the data reuse for intra-layer assignments, like output-stationary in ShiDianNao~\cite{shidiannao} and Envision~\cite{envision}, weight-stationary in NeuFlow~\cite{neuflow} and Nvdla~\cite{nvdla}, input-stationary in SCNN~\cite{scnn}, and row-stationary in Eyeriss~\cite{eyeriss}.
Based on these primitive dataflow patterns, extensive studies explored the optimal tiling and reordering schemes via brute-force, feedback-based, and constraint optimization approaches~\cite{timeloop, GAMMA, CoSA}.
These works focus on layer-level optimization, missing the graph information at a higher level.
The efficiency of tile updates depends on the memory architecture.
Simba~\cite{simba, simba-vlsi} and NN-Baton~\cite{nnbaton} view each tile as an independent workload so that the tile size has a prominent impact on memory access due to halo regions.
Motivated by traditional vision processors, Ascend~\cite{Ascend1} and DRQ~\cite{DRQ} employ line buffers to achieve data reuse in the row direction, but the line buffer cannot well support the 2D-tiling reuse in both row and column directions.

\subsection{Inter-layer Optimization}

Intra-layer scheduling is sub-optimal, which is limited by the data reuse within a layer.
Therefore, Fused-CNN~\cite{fused-CNN}, SR-CNN~\cite{fused-cnn2}, and LCP~\cite{lcp} introduce layer fusion method that cache intermediate data on-chip to reduce data transfer overhead using handcrafted or heuristic methods for fusion partition.
Although Irregular-NN~\cite{irregular} suggests a customized-DP algorithm, the exploration space is constrained because the layers in an assignment need to be successive in a specific order.
A recent work named DNNFuser~\cite{DNNFuser} employs an RL-based method, but their formulation towards 1D layer-fusion is hard to handle complex irregular networks.
Tangram~\cite{tangram} and Atomic~\cite{atomic} schedule DNN workloads on a multi-core (scalable) accelerator, but they focus on executing a single layer on each core at a time rather than processing multiple layers with local data reuse. Also, some previous works~\cite{dnn-partitioning, Placeto, Spotlight} tackle the workload placement problem for multiple devides without discussing the downstream execution on each device.

Cocco proposes an automatic framework for inter-layer scheduling with a comprehensive memory scheme.
It focuses on the fundamental core-level temporal execution that can be potentially scaled up to the multi-core or multi-device scenario with a spatial parallelism mechanism.

\subsection{Design-Space Exploration for Memory}

Memory design exploration methods lie primarily on two sides: analysis-driven and search-driven.
For the analysis-driven method, Chen~\textit{et al.}~\cite{lower-bound} leverage red-blue pebble models to derive the proper memory capacity representations.
Subsequently, Cai~\textit{et al.}~\cite{olympus} propose Olympus, which generalizes a framework to a batch of successive layers and also fills up with more scheduling and data reuse techniques.
However, they are difficult to represent a subgraph with complex inter-layer connections.
As for the search-driven method, Xiao~\textit{et al.}~\cite{hasco}, Kwon~\textit{et al.}~\cite{understand}, and Feng~\textit{et al.}~\cite{ERDSE} explore the memory configuration for the layer-level assignment using the brute-force search, while Kao~\textit{et al.}~\cite{digamma} employ a genetic algorithm to improve the efficiency.
These works principally focus on the layer-level information, while in comparison, Cocco exploits graph-level features for the better optimization.
\section{Conclusion}

While layer-level scheduling is widely studied to improve memory efficiency, graph-level optimization remains relatively unexplored.
This paper proposed a graph-level dataflow with the corresponding memory management scheme that enables flexible graph partitions with high memory utilization.
On top of it, we propose Cocco, a framework to provide a recommended memory configuration with graph-level scheduling strategies.
Cocco shows outstanding graph partition ability compared to the greedy algorithm and DP employed in previous works and enables efficient graph-level hardware-mapping co-exploration.
\hl{This paper helps to provide an implementation philosophy for the accelerator memory and better deployment for it.}

%%
%% The acknowledgments section is defined using the "acks" environment
%% (and NOT an unnumbered section). This ensures the proper
%% identification of the section in the article metadata, and the
%% consistent spelling of the heading.
\begin{acks}
This research was partially supported by National Key R\&D Program of China (2022YFB2804103), Tsinghua University Dushi Program, and Tsinghua University Talent Program.
We would like to appreciate all the anonymous reviewers for their valuable feedback.
\end{acks}

%%
%% The next two lines define the bibliography style to be used, and
%% the bibliography file.
\bibliographystyle{ACM-Reference-Format}
\bibliography{references}

%%% -*-BibTeX-*-
%%% Do NOT edit. File created by BibTeX with style
%%% ACM-Reference-Format-Journals [18-Jan-2012].

\begin{thebibliography}{75}

%%% ====================================================================
%%% NOTE TO THE USER: you can override these defaults by providing
%%% customized versions of any of these macros before the \bibliography
%%% command.  Each of them MUST provide its own final punctuation,
%%% except for \shownote{}, \showDOI{}, and \showURL{}.  The latter two
%%% do not use final punctuation, in order to avoid confusing it with
%%% the Web address.
%%%
%%% To suppress output of a particular field, define its macro to expand
%%% to an empty string, or better, \unskip, like this:
%%%
%%% \newcommand{\showDOI}[1]{\unskip}   % LaTeX syntax
%%%
%%% \def \showDOI #1{\unskip}           % plain TeX syntax
%%%
%%% ====================================================================

\ifx \showCODEN    \undefined \def \showCODEN     #1{\unskip}     \fi
\ifx \showDOI      \undefined \def \showDOI       #1{#1}\fi
\ifx \showISBNx    \undefined \def \showISBNx     #1{\unskip}     \fi
\ifx \showISBNxiii \undefined \def \showISBNxiii  #1{\unskip}     \fi
\ifx \showISSN     \undefined \def \showISSN      #1{\unskip}     \fi
\ifx \showLCCN     \undefined \def \showLCCN      #1{\unskip}     \fi
\ifx \shownote     \undefined \def \shownote      #1{#1}          \fi
\ifx \showarticletitle \undefined \def \showarticletitle #1{#1}   \fi
\ifx \showURL      \undefined \def \showURL       {\relax}        \fi
% The following commands are used for tagged output and should be
% invisible to TeX
\providecommand\bibfield[2]{#2}
\providecommand\bibinfo[2]{#2}
\providecommand\natexlab[1]{#1}
\providecommand\showeprint[2][]{arXiv:#2}

\bibitem[Abts et~al\mbox{.}(2020)]%
        {Groq}
\bibfield{author}{\bibinfo{person}{Dennis Abts}, \bibinfo{person}{Jonathan Ross}, \bibinfo{person}{Jonathan Sparling}, \bibinfo{person}{Mark Wong{-}VanHaren}, \bibinfo{person}{Max Baker}, \bibinfo{person}{Tom Hawkins}, \bibinfo{person}{Andrew Bell}, \bibinfo{person}{John Thompson}, \bibinfo{person}{Temesghen Kahsai}, \bibinfo{person}{Garrin Kimmell}, \bibinfo{person}{Jennifer Hwang}, \bibinfo{person}{Rebekah Leslie{-}Hurd}, \bibinfo{person}{Michael Bye}, \bibinfo{person}{E.~R. Creswick}, \bibinfo{person}{Matthew Boyd}, \bibinfo{person}{Mahitha Venigalla}, \bibinfo{person}{Evan Laforge}, \bibinfo{person}{Jon Purdy}, \bibinfo{person}{Purushotham Kamath}, \bibinfo{person}{Dinesh Maheshwari}, \bibinfo{person}{Michael Beidler}, \bibinfo{person}{Geert Rosseel}, \bibinfo{person}{Omar Ahmad}, \bibinfo{person}{Gleb Gagarin}, \bibinfo{person}{Richard Czekalski}, \bibinfo{person}{Ashay Rane}, \bibinfo{person}{Sahil Parmar}, \bibinfo{person}{Jeff Werner}, \bibinfo{person}{Jim Sproch}, \bibinfo{person}{Adrian Macias},
  {and} \bibinfo{person}{Brian Kurtz}.} \bibinfo{year}{2020}\natexlab{}.
\newblock \showarticletitle{Think Fast: {A} Tensor Streaming Processor {(TSP)} for Accelerating Deep Learning Workloads}. In \bibinfo{booktitle}{\emph{Proceedings of the 47th {ACM/IEEE} Annual International Symposium on Computer Architecture ({ISCA})}}. \bibinfo{publisher}{{IEEE}}, \bibinfo{address}{Valencia, Spain}, \bibinfo{pages}{145--158}.
\newblock


\bibitem[Addanki et~al\mbox{.}(2019)]%
        {Placeto}
\bibfield{author}{\bibinfo{person}{Ravichandra Addanki}, \bibinfo{person}{Shaileshh~Bojja Venkatakrishnan}, \bibinfo{person}{Shreyan Gupta}, \bibinfo{person}{Hongzi Mao}, {and} \bibinfo{person}{Mohammad Alizadeh}.} \bibinfo{year}{2019}\natexlab{}.
\newblock \showarticletitle{Learning Generalizable Device Placement Algorithms for Distributed Machine Learning}. In \bibinfo{booktitle}{\emph{Advances in Neural Information Processing Systems ({NeurIPS})}}, \bibfield{editor}{\bibinfo{person}{Hanna~M. Wallach}, \bibinfo{person}{Hugo Larochelle}, \bibinfo{person}{Alina Beygelzimer}, \bibinfo{person}{Florence d'Alch{\'{e}}{-}Buc}, \bibinfo{person}{Emily~B. Fox}, {and} \bibinfo{person}{Roman Garnett}} (Eds.). \bibinfo{publisher}{OpenReview.net}, \bibinfo{address}{Vancouver, BC, Canada}, \bibinfo{pages}{3983--3993}.
\newblock


\bibitem[Ahn et~al\mbox{.}(2020)]%
        {chaos}
\bibfield{author}{\bibinfo{person}{Byung~Hoon Ahn}, \bibinfo{person}{Jinwon Lee}, \bibinfo{person}{Jamie~Menjay Lin}, \bibinfo{person}{Hsin{-}Pai Cheng}, \bibinfo{person}{Jilei Hou}, {and} \bibinfo{person}{Hadi Esmaeilzadeh}.} \bibinfo{year}{2020}\natexlab{}.
\newblock \showarticletitle{Ordering Chaos: Memory-Aware Scheduling of Irregularly Wired Neural Networks for Edge Devices}. In \bibinfo{booktitle}{\emph{Proceedings of Machine Learning and Systems ({MLSys})}}, \bibfield{editor}{\bibinfo{person}{Inderjit~S. Dhillon}, \bibinfo{person}{Dimitris~S. Papailiopoulos}, {and} \bibinfo{person}{Vivienne Sze}} (Eds.). \bibinfo{publisher}{mlsys.org}, \bibinfo{address}{Austin, TX, USA}, \bibinfo{pages}{1--14}.
\newblock


\bibitem[Alwani et~al\mbox{.}(2016)]%
        {fused-CNN}
\bibfield{author}{\bibinfo{person}{Manoj Alwani}, \bibinfo{person}{Han Chen}, \bibinfo{person}{Michael Ferdman}, {and} \bibinfo{person}{Peter~A. Milder}.} \bibinfo{year}{2016}\natexlab{}.
\newblock \showarticletitle{Fused-layer {CNN} accelerators}. In \bibinfo{booktitle}{\emph{Proceedings of the 49th {IEEE/ACM} International Symposium on Microarchitecture ({MICRO})}}. \bibinfo{publisher}{{IEEE} Computer Society}, \bibinfo{address}{Taipei, Taiwan}, \bibinfo{pages}{22:1--22:12}.
\newblock


\bibitem[Arteries(2022)]%
        {arteries}
\bibfield{author}{\bibinfo{person}{Arteries}.} \bibinfo{year}{2022}\natexlab{}.
\newblock \bibinfo{title}{Arteries IP Homepage}.
\newblock \bibinfo{howpublished}{\url{https://www.arteris.com}}.
\newblock


\bibitem[Bajic and Vasiljevic(2020)]%
        {Grayskull2}
\bibfield{author}{\bibinfo{person}{Ljubisa Bajic} {and} \bibinfo{person}{Jasmina Vasiljevic}.} \bibinfo{year}{2020}\natexlab{}.
\newblock \showarticletitle{Compute substrate for Software 2.0}. In \bibinfo{booktitle}{\emph{Proceedings of the {IEEE} Hot Chips 32 Symposium ({HCS})}}. \bibinfo{publisher}{{IEEE}}, \bibinfo{address}{Palo Alto, CA, USA}, \bibinfo{pages}{1--31}.
\newblock


\bibitem[Bannon et~al\mbox{.}(2019)]%
        {FSD}
\bibfield{author}{\bibinfo{person}{Pete Bannon}, \bibinfo{person}{Ganesh Venkataramanan}, \bibinfo{person}{Debjit~Das Sarma}, {and} \bibinfo{person}{Emil Talpes}.} \bibinfo{year}{2019}\natexlab{}.
\newblock \showarticletitle{Computer and Redundancy Solution for the Full Self-Driving Computer}. In \bibinfo{booktitle}{\emph{Proceedings of the {IEEE} Hot Chips 31 Symposium (HCS)}}. \bibinfo{publisher}{{IEEE}}, \bibinfo{address}{Cupertino, CA, USA}, \bibinfo{pages}{1--22}.
\newblock


\bibitem[Burgess(2019)]%
        {T4}
\bibfield{author}{\bibinfo{person}{John Burgess}.} \bibinfo{year}{2019}\natexlab{}.
\newblock \showarticletitle{{RTX} {ON} - The {NVIDIA} {TURING} {GPU}}. In \bibinfo{booktitle}{\emph{Proceedings of the {IEEE} Hot Chips 31 Symposium (HCS)}}. \bibinfo{publisher}{{IEEE}}, \bibinfo{address}{Cupertino, CA, USA}, \bibinfo{pages}{1--27}.
\newblock


\bibitem[Cai et~al\mbox{.}(2022)]%
        {olympus}
\bibfield{author}{\bibinfo{person}{Xuyi Cai}, \bibinfo{person}{Ying Wang}, \bibinfo{person}{Kaijie Tu}, \bibinfo{person}{Chengsi Gao}, {and} \bibinfo{person}{Lei Zhang}.} \bibinfo{year}{2022}\natexlab{}.
\newblock \showarticletitle{Olympus: Reaching Memory-Optimality on DNN Processors}.
\newblock \bibinfo{journal}{\emph{IEEE Transactions on Computers ({TC})}} \bibinfo{volume}{71}, \bibinfo{number}{8} (\bibinfo{year}{2022}), \bibinfo{pages}{1939--1951}.
\newblock


\bibitem[Chatarasi et~al\mbox{.}(2022)]%
        {marvel}
\bibfield{author}{\bibinfo{person}{Prasanth Chatarasi}, \bibinfo{person}{Hyoukjun Kwon}, \bibinfo{person}{Angshuman Parashar}, \bibinfo{person}{Michael Pellauer}, \bibinfo{person}{Tushar Krishna}, {and} \bibinfo{person}{Vivek Sarkar}.} \bibinfo{year}{2022}\natexlab{}.
\newblock \showarticletitle{Marvel: {A} Data-Centric Approach for Mapping Deep Learning Operators on Spatial Accelerators}.
\newblock \bibinfo{journal}{\emph{{ACM} Transactions on Architecture and Code Optimization}} \bibinfo{volume}{19}, \bibinfo{number}{1} (\bibinfo{year}{2022}), \bibinfo{pages}{6:1--6:26}.
\newblock


\bibitem[Chatha(2021)]%
        {Qualcomm}
\bibfield{author}{\bibinfo{person}{Karam Chatha}.} \bibinfo{year}{2021}\natexlab{}.
\newblock \showarticletitle{Qualcomm\({}^{\mbox{{\textregistered}}}\) Cloud {Al}-100: 12{TOPS/W} Scalable, High Performance and Low Latency Deep Learning Inference Accelerator}. In \bibinfo{booktitle}{\emph{Proceedings of the {IEEE} Hot Chips 33 Symposium ({HCS})}}. \bibinfo{publisher}{{IEEE}}, \bibinfo{address}{Palo Alto, CA, USA}, \bibinfo{pages}{1--19}.
\newblock


\bibitem[Chen et~al\mbox{.}(2020)]%
        {lower-bound}
\bibfield{author}{\bibinfo{person}{Xiaoming Chen}, \bibinfo{person}{Yinhe Han}, {and} \bibinfo{person}{Yu Wang}.} \bibinfo{year}{2020}\natexlab{}.
\newblock \showarticletitle{Communication Lower Bound in Convolution Accelerators}. In \bibinfo{booktitle}{\emph{Proceedings of the {IEEE} International Symposium on High Performance Computer Architecture ({HPCA})}}. \bibinfo{publisher}{{IEEE}}, \bibinfo{address}{San Diego, CA, USA}, \bibinfo{pages}{529--541}.
\newblock


\bibitem[Chen et~al\mbox{.}(2016)]%
        {eyeriss}
\bibfield{author}{\bibinfo{person}{Yu{-}Hsin Chen}, \bibinfo{person}{Joel~S. Emer}, {and} \bibinfo{person}{Vivienne Sze}.} \bibinfo{year}{2016}\natexlab{}.
\newblock \showarticletitle{Eyeriss: {A} Spatial Architecture for Energy-Efficient Dataflow for Convolutional Neural Networks}. In \bibinfo{booktitle}{\emph{Proceedings of the {ACM/IEEE} Annual International Symposium on Computer Architecture ({ISCA})}}. \bibinfo{publisher}{{IEEE} Computer Society}, \bibinfo{address}{Seoul, South Korea}, \bibinfo{pages}{367--379}.
\newblock


\bibitem[Du et~al\mbox{.}(2015)]%
        {shidiannao}
\bibfield{author}{\bibinfo{person}{Zidong Du}, \bibinfo{person}{Robert Fasthuber}, \bibinfo{person}{Tianshi Chen}, \bibinfo{person}{Paolo Ienne}, \bibinfo{person}{Ling Li}, \bibinfo{person}{Tao Luo}, \bibinfo{person}{Xiaobing Feng}, \bibinfo{person}{Yunji Chen}, {and} \bibinfo{person}{Olivier Temam}.} \bibinfo{year}{2015}\natexlab{}.
\newblock \showarticletitle{ShiDianNao: shifting vision processing closer to the sensor}. In \bibinfo{booktitle}{\emph{Proceedings of the {ACM/IEEE} Annual International Symposium on Computer Architecture ({ISCA})}}. \bibinfo{publisher}{{ACM}}, \bibinfo{address}{Portland, OR, USA}, \bibinfo{pages}{92--104}.
\newblock


\bibitem[Farabet et~al\mbox{.}(2011)]%
        {neuflow}
\bibfield{author}{\bibinfo{person}{Cl{\'{e}}ment Farabet}, \bibinfo{person}{Berin Martini}, \bibinfo{person}{B. Corda}, \bibinfo{person}{Polina Akselrod}, \bibinfo{person}{Eugenio Culurciello}, {and} \bibinfo{person}{Yann LeCun}.} \bibinfo{year}{2011}\natexlab{}.
\newblock \showarticletitle{NeuFlow: {A} runtime reconfigurable dataflow processor for vision}. In \bibinfo{booktitle}{\emph{Proceedings of the {IEEE} Conference on Computer Vision and Pattern Recognition (CVPR) Workshops}}. \bibinfo{publisher}{{IEEE} Computer Society}, \bibinfo{address}{Colorado Springs, CO, USA}, \bibinfo{pages}{109--116}.
\newblock


\bibitem[Feng et~al\mbox{.}(2021)]%
        {ERDSE}
\bibfield{author}{\bibinfo{person}{Kaijie Feng}, \bibinfo{person}{Xiaoya Fan}, \bibinfo{person}{Jianfeng An}, \bibinfo{person}{Xiping Wang}, \bibinfo{person}{Kaiyue Di}, \bibinfo{person}{Jiangfei Li}, \bibinfo{person}{Minghao Lu}, {and} \bibinfo{person}{Chuxi Li}.} \bibinfo{year}{2021}\natexlab{}.
\newblock \showarticletitle{{ERDSE:} efficient reinforcement learning based design space exploration method for {CNN} accelerator on resource limited platform}.
\newblock \bibinfo{journal}{\emph{Graphics and Visual Computing}}  \bibinfo{volume}{4} (\bibinfo{year}{2021}), \bibinfo{pages}{1--11}.
\newblock


\bibitem[Funahashi(1989)]%
        {mlp2}
\bibfield{author}{\bibinfo{person}{Ken{-}ichi Funahashi}.} \bibinfo{year}{1989}\natexlab{}.
\newblock \showarticletitle{On the approximate realization of continuous mappings by neural networks}.
\newblock \bibinfo{journal}{\emph{Neural Networks}} \bibinfo{volume}{2}, \bibinfo{number}{3} (\bibinfo{year}{1989}), \bibinfo{pages}{183--192}.
\newblock


\bibitem[Gao et~al\mbox{.}(2019)]%
        {tangram}
\bibfield{author}{\bibinfo{person}{Mingyu Gao}, \bibinfo{person}{Xuan Yang}, \bibinfo{person}{Jing Pu}, \bibinfo{person}{Mark Horowitz}, {and} \bibinfo{person}{Christos Kozyrakis}.} \bibinfo{year}{2019}\natexlab{}.
\newblock \showarticletitle{{TANGRAM:} Optimized Coarse-Grained Dataflow for Scalable {NN} Accelerators}. In \bibinfo{booktitle}{\emph{Proceedings of the International Conference on Architectural Support for Programming Languages and Operating Systems ({ASPLOS})}}. \bibinfo{publisher}{{ACM}}, \bibinfo{address}{Providence, RI, USA}, \bibinfo{pages}{807--820}.
\newblock


\bibitem[Gao et~al\mbox{.}(2018)]%
        {Spotlight}
\bibfield{author}{\bibinfo{person}{Yuanxiang Gao}, \bibinfo{person}{Li Chen}, {and} \bibinfo{person}{Baochun Li}.} \bibinfo{year}{2018}\natexlab{}.
\newblock \showarticletitle{Spotlight: Optimizing Device Placement for Training Deep Neural Networks}. In \bibinfo{booktitle}{\emph{Proceedings of the 35th International Conference on Machine Learning {(ICML)}}} \emph{(\bibinfo{series}{Proceedings of Machine Learning Research}, Vol.~\bibinfo{volume}{80})}, \bibfield{editor}{\bibinfo{person}{Jennifer~G. Dy} {and} \bibinfo{person}{Andreas Krause}} (Eds.). \bibinfo{publisher}{{PMLR}}, \bibinfo{address}{Stockholm, Sweden}, \bibinfo{pages}{1662--1670}.
\newblock


\bibitem[He et~al\mbox{.}(2016)]%
        {res}
\bibfield{author}{\bibinfo{person}{Kaiming He}, \bibinfo{person}{Xiangyu Zhang}, \bibinfo{person}{Shaoqing Ren}, {and} \bibinfo{person}{Sun Jian}.} \bibinfo{year}{2016}\natexlab{}.
\newblock \showarticletitle{Deep Residual Learning for Image Recognition}. In \bibinfo{booktitle}{\emph{Proceedings of the {IEEE} Conference on Computer Vision and Pattern Recognition ({CVPR})}}. \bibinfo{publisher}{{IEEE} Computer Society}, \bibinfo{address}{Las Vegas, NV, USA}, \bibinfo{pages}{770--778}.
\newblock


\bibitem[Hegde et~al\mbox{.}(2021)]%
        {mindmap}
\bibfield{author}{\bibinfo{person}{Kartik Hegde}, \bibinfo{person}{Po{-}An Tsai}, \bibinfo{person}{Sitao Huang}, \bibinfo{person}{Vikas Chandra}, \bibinfo{person}{Angshuman Parashar}, {and} \bibinfo{person}{Christopher~W. Fletcher}.} \bibinfo{year}{2021}\natexlab{}.
\newblock \showarticletitle{Mind mappings: enabling efficient algorithm-accelerator mapping space search}. In \bibinfo{booktitle}{\emph{Proceedings of the 26th {ACM} International Conference on Architectural Support for Programming Languages and Operating Systems ({ASPLOS})}}, \bibfield{editor}{\bibinfo{person}{Tim Sherwood}, \bibinfo{person}{Emery~D. Berger}, {and} \bibinfo{person}{Christos Kozyrakis}} (Eds.). \bibinfo{publisher}{{ACM}}, \bibinfo{address}{Virtual Event, USA}, \bibinfo{pages}{943--958}.
\newblock


\bibitem[Hornik et~al\mbox{.}(1989)]%
        {mlp3}
\bibfield{author}{\bibinfo{person}{Kurt Hornik}, \bibinfo{person}{Maxwell~B. Stinchcombe}, {and} \bibinfo{person}{Halbert White}.} \bibinfo{year}{1989}\natexlab{}.
\newblock \showarticletitle{Multilayer feedforward networks are universal approximators}.
\newblock \bibinfo{journal}{\emph{Neural Networks}} \bibinfo{volume}{2}, \bibinfo{number}{5} (\bibinfo{year}{1989}), \bibinfo{pages}{359--366}.
\newblock


\bibitem[Huang et~al\mbox{.}(2021)]%
        {CoSA}
\bibfield{author}{\bibinfo{person}{Qijing Huang}, \bibinfo{person}{Aravind Kalaiah}, \bibinfo{person}{Minwoo Kang}, \bibinfo{person}{James Demmel}, \bibinfo{person}{Grace Dinh}, \bibinfo{person}{John Wawrzynek}, \bibinfo{person}{Thomas Norell}, {and} \bibinfo{person}{Yakun~Sophia Shao}.} \bibinfo{year}{2021}\natexlab{}.
\newblock \showarticletitle{CoSA: Scheduling by Constrained Optimization for Spatial Accelerators}. In \bibinfo{booktitle}{\emph{Proceedings of the {ACM/IEEE} Annual International Symposium on Computer Architecture ({ISCA})}}. \bibinfo{publisher}{{IEEE}}, \bibinfo{address}{Valencia, Spain}, \bibinfo{pages}{554--566}.
\newblock


\bibitem[Ignjatovic et~al\mbox{.}(2022)]%
        {Wormhole}
\bibfield{author}{\bibinfo{person}{Drago Ignjatovic}, \bibinfo{person}{Daniel~W. Bailey}, {and} \bibinfo{person}{Ljubisa Bajic}.} \bibinfo{year}{2022}\natexlab{}.
\newblock \showarticletitle{The Wormhole {AI} Training Processor}. In \bibinfo{booktitle}{\emph{Proceedings of the {IEEE} International Solid-State Circuits Conference ({ISSCC})}}. \bibinfo{publisher}{{IEEE}}, \bibinfo{address}{San Francisco, CA, USA}, \bibinfo{pages}{356--358}.
\newblock


\bibitem[Jangda and Bondhugula(2018)]%
        {DBLP:conf/ppopp/JangdaB18}
\bibfield{author}{\bibinfo{person}{Abhinav Jangda} {and} \bibinfo{person}{Uday Bondhugula}.} \bibinfo{year}{2018}\natexlab{}.
\newblock \showarticletitle{An effective fusion and tile size model for optimizing image processing pipelines}. In \bibinfo{booktitle}{\emph{Proceedings of the 23rd {ACM} {SIGPLAN} Symposium on Principles and Practice of Parallel Programming {(PPoPP)}}}, \bibfield{editor}{\bibinfo{person}{Andreas Krall} {and} \bibinfo{person}{Thomas~R. Gross}} (Eds.). \bibinfo{publisher}{{ACM}}, \bibinfo{address}{Vienna, Austria}, \bibinfo{pages}{261--275}.
\newblock


\bibitem[Jiao et~al\mbox{.}(2020b)]%
        {Hanguang1}
\bibfield{author}{\bibinfo{person}{Yang Jiao}, \bibinfo{person}{Liang Han}, \bibinfo{person}{Rong Jin}, \bibinfo{person}{Yi{-}Jung Su}, \bibinfo{person}{Chiente Ho}, \bibinfo{person}{Li Yin}, \bibinfo{person}{Yun Li}, \bibinfo{person}{Long Chen}, \bibinfo{person}{Zhen Chen}, \bibinfo{person}{Lu Liu}, \bibinfo{person}{Zhuyu He}, \bibinfo{person}{Yu Yan}, \bibinfo{person}{Jun He}, \bibinfo{person}{Jun Mao}, \bibinfo{person}{Xiaotao Zai}, \bibinfo{person}{Xuejun Wu}, \bibinfo{person}{Yongquan Zhou}, \bibinfo{person}{Mingqiu Gu}, \bibinfo{person}{Guocai Zhu}, \bibinfo{person}{Rong Zhong}, \bibinfo{person}{Wenyuan Lee}, \bibinfo{person}{Ping Chen}, \bibinfo{person}{Yiping Chen}, \bibinfo{person}{Weiliang Li}, \bibinfo{person}{Deyu Xiao}, \bibinfo{person}{Qing Yan}, \bibinfo{person}{Mingyuan Zhuang}, \bibinfo{person}{Jiejun Chen}, \bibinfo{person}{Yun Tian}, \bibinfo{person}{Yingzi Lin}, \bibinfo{person}{Wei Wu}, \bibinfo{person}{Hao Li}, {and} \bibinfo{person}{Zesheng Dou}.} \bibinfo{year}{2020}\natexlab{b}.
\newblock \showarticletitle{{A} 12nm Programmable Convolution-Efficient Neural-Processing-Unit Chip Achieving 825TOPS}. In \bibinfo{booktitle}{\emph{Proceedings of the {IEEE} International Solid-State Circuits Conference ({ISSCC})}}. \bibinfo{publisher}{{IEEE}}, \bibinfo{address}{San Francisco, CA, USA}, \bibinfo{pages}{136--140}.
\newblock


\bibitem[Jiao et~al\mbox{.}(2020a)]%
        {Hanguang2}
\bibfield{author}{\bibinfo{person}{Yang Jiao}, \bibinfo{person}{Liang Han}, {and} \bibinfo{person}{Xin Long}.} \bibinfo{year}{2020}\natexlab{a}.
\newblock \showarticletitle{Hanguang 800 {NPU} - The Ultimate {AI} Inference Solution for Data Centers}. In \bibinfo{booktitle}{\emph{Proceedings of the {IEEE} Hot Chips 32 Symposium ({HCS})}}. \bibinfo{publisher}{{IEEE}}, \bibinfo{address}{Palo Alto, CA, USA}, \bibinfo{pages}{1--29}.
\newblock


\bibitem[Jouppi et~al\mbox{.}(2021)]%
        {TPUv4i}
\bibfield{author}{\bibinfo{person}{Norman~P. Jouppi}, \bibinfo{person}{Doe~Hyun Yoon}, \bibinfo{person}{Matthew Ashcraft}, \bibinfo{person}{Mark Gottscho}, \bibinfo{person}{Thomas~B. Jablin}, \bibinfo{person}{George Kurian}, \bibinfo{person}{James Laudon}, \bibinfo{person}{Sheng Li}, \bibinfo{person}{Peter~C. Ma}, \bibinfo{person}{Xiaoyu Ma}, \bibinfo{person}{Thomas Norrie}, \bibinfo{person}{Nishant Patil}, \bibinfo{person}{Sushma Prasad}, \bibinfo{person}{Cliff Young}, \bibinfo{person}{Zongwei Zhou}, {and} \bibinfo{person}{David~A. Patterson}.} \bibinfo{year}{2021}\natexlab{}.
\newblock \showarticletitle{Ten Lessons From Three Generations Shaped Google's TPUv4i : Industrial Product}. In \bibinfo{booktitle}{\emph{Proceedings of the 48th {ACM/IEEE} Annual International Symposium on Computer Architecture ({ISCA})}}. \bibinfo{publisher}{{IEEE}}, \bibinfo{address}{Valencia, Spain}, \bibinfo{pages}{1--14}.
\newblock


\bibitem[Kao et~al\mbox{.}(2022a)]%
        {DNNFuser}
\bibfield{author}{\bibinfo{person}{Sheng{-}Chun Kao}, \bibinfo{person}{Xiaoyu Huang}, {and} \bibinfo{person}{Tushar Krishna}.} \bibinfo{year}{2022}\natexlab{a}.
\newblock \showarticletitle{DNNFuser: Generative Pre-Trained Transformer as a Generalized Mapper for Layer Fusion in {DNN} Accelerators}.
\newblock \bibinfo{journal}{\emph{arXiv preprint arXiv:2201.11218}}  \bibinfo{volume}{abs/2201.11218} (\bibinfo{year}{2022}), \bibinfo{pages}{1--8}.
\newblock


\bibitem[Kao and Krishna(2020)]%
        {GAMMA}
\bibfield{author}{\bibinfo{person}{Sheng{-}Chun Kao} {and} \bibinfo{person}{Tushar Krishna}.} \bibinfo{year}{2020}\natexlab{}.
\newblock \showarticletitle{{GAMMA:} Automating the {HW} Mapping of {DNN} Models on Accelerators via Genetic Algorithm}. In \bibinfo{booktitle}{\emph{Proceedings of the {IEEE/ACM} International Conference On Computer Aided Design ({ICCAD})}}. \bibinfo{publisher}{{IEEE}}, \bibinfo{address}{San Diego, CA, USA}, \bibinfo{pages}{44:1--44:9}.
\newblock


\bibitem[Kao and Krishna(2022)]%
        {magma}
\bibfield{author}{\bibinfo{person}{Sheng{-}Chun Kao} {and} \bibinfo{person}{Tushar Krishna}.} \bibinfo{year}{2022}\natexlab{}.
\newblock \showarticletitle{{MAGMA:} An Optimization Framework for Mapping Multiple DNNs on Multiple Accelerator Cores}. In \bibinfo{booktitle}{\emph{{IEEE} International Symposium on High-Performance Computer Architecture, ({HPCA})}}. \bibinfo{publisher}{{IEEE}}, \bibinfo{address}{Seoul, South Korea}, \bibinfo{pages}{814--830}.
\newblock


\bibitem[Kao et~al\mbox{.}(2022b)]%
        {digamma}
\bibfield{author}{\bibinfo{person}{Sheng{-}Chun Kao}, \bibinfo{person}{Michael Pellauer}, \bibinfo{person}{Angshuman Parashar}, {and} \bibinfo{person}{Tushar Krishna}.} \bibinfo{year}{2022}\natexlab{b}.
\newblock \showarticletitle{DiGamma: Domain-aware Genetic Algorithm for HW-Mapping Co-optimization for {DNN} Accelerators}. In \bibinfo{booktitle}{\emph{Proceedings of the Design, Automation {\&} Test in Europe Conference {\&} Exhibition ({DATE})}}, \bibfield{editor}{\bibinfo{person}{Cristiana Bolchini}, \bibinfo{person}{Ingrid Verbauwhede}, {and} \bibinfo{person}{Ioana Vatajelu}} (Eds.). \bibinfo{publisher}{{IEEE}}, \bibinfo{address}{Antwerp, Belgium}, \bibinfo{pages}{232--237}.
\newblock


\bibitem[Kirkpatrick et~al\mbox{.}(1983)]%
        {SA}
\bibfield{author}{\bibinfo{person}{Scott Kirkpatrick}, \bibinfo{person}{D.~Gelatt Jr.}, {and} \bibinfo{person}{Mario~P. Vecchi}.} \bibinfo{year}{1983}\natexlab{}.
\newblock \showarticletitle{Optimization by Simmulated Annealing}.
\newblock \bibinfo{journal}{\emph{Sci.}} \bibinfo{volume}{220}, \bibinfo{number}{4598} (\bibinfo{year}{1983}), \bibinfo{pages}{671--680}.
\newblock


\bibitem[Knowles(2017)]%
        {IPUv1}
\bibfield{author}{\bibinfo{person}{Simon Knowles}.} \bibinfo{year}{2017}\natexlab{}.
\newblock \showarticletitle{Scalable Silicon Compute}. In \bibinfo{booktitle}{\emph{Workshop on Deep Learning At Supercomputer Scale, NIPS}}. \bibinfo{publisher}{OpenReview.net}, \bibinfo{address}{Long Beach, CA, USA}, \bibinfo{pages}{1--22}.
\newblock


\bibitem[Knowles(2021)]%
        {IPUv2}
\bibfield{author}{\bibinfo{person}{Simon Knowles}.} \bibinfo{year}{2021}\natexlab{}.
\newblock \showarticletitle{Graphcore}. In \bibinfo{booktitle}{\emph{Proceedings of the {IEEE} Hot Chips 33 Symposium ({HCS})}}. \bibinfo{publisher}{{IEEE}}, \bibinfo{address}{Palo Alto, CA, USA}, \bibinfo{pages}{1--25}.
\newblock


\bibitem[Krizhevsky et~al\mbox{.}(2012)]%
        {alexNet}
\bibfield{author}{\bibinfo{person}{Alex Krizhevsky}, \bibinfo{person}{Ilya Sutskever}, {and} \bibinfo{person}{Geoffrey~E. Hinton}.} \bibinfo{year}{2012}\natexlab{}.
\newblock \showarticletitle{ImageNet Classification with Deep Convolutional Neural Networks}. In \bibinfo{booktitle}{\emph{Proceedings of the 26th Annual Conference on Neural Information Processing Systems ({NIPS})}}. \bibinfo{publisher}{Curran Associates, Inc.}, \bibinfo{address}{Lake Tahoe, Nevada, United States}, \bibinfo{pages}{1106--1114}.
\newblock


\bibitem[Kwon et~al\mbox{.}(2019)]%
        {understand}
\bibfield{author}{\bibinfo{person}{Hyoukjun Kwon}, \bibinfo{person}{Prasanth Chatarasi}, \bibinfo{person}{Michael Pellauer}, \bibinfo{person}{Angshuman Parashar}, \bibinfo{person}{Vivek Sarkar}, {and} \bibinfo{person}{Tushar Krishna}.} \bibinfo{year}{2019}\natexlab{}.
\newblock \showarticletitle{Understanding Reuse, Performance, and Hardware Cost of {DNN} Dataflow: {A} Data-Centric Approach}. In \bibinfo{booktitle}{\emph{Proceedings of the {IEEE/ACM} International Symposium on Microarchitecture ({MICRO})}}. \bibinfo{publisher}{{ACM}}, \bibinfo{address}{Columbus, OH, USA}, \bibinfo{pages}{754--768}.
\newblock


\bibitem[Lee et~al\mbox{.}(2019)]%
        {fused-cnn2}
\bibfield{author}{\bibinfo{person}{Juhyoung Lee}, \bibinfo{person}{Dongjoo Shin}, \bibinfo{person}{Jinsu Lee}, \bibinfo{person}{Jinmook Lee}, \bibinfo{person}{Sanghoon Kang}, {and} \bibinfo{person}{Hoi{-}Jun Yoo}.} \bibinfo{year}{2019}\natexlab{}.
\newblock \showarticletitle{A Full {HD} 60 fps {CNN} Super Resolution Processor with Selective Caching based Layer Fusion for Mobile Devices}. In \bibinfo{booktitle}{\emph{Proceedings of the Symposium on {VLSI} Circuits}}. \bibinfo{publisher}{{IEEE}}, \bibinfo{address}{Kyoto, Japan}, \bibinfo{pages}{302--303}.
\newblock


\bibitem[Lewicki and Marino(2004)]%
        {mlp1}
\bibfield{author}{\bibinfo{person}{Grzegorz Lewicki} {and} \bibinfo{person}{Giuseppe Marino}.} \bibinfo{year}{2004}\natexlab{}.
\newblock \showarticletitle{Approximation of functions of finite variation by superpositions of a Sigmoidal function}.
\newblock \bibinfo{journal}{\emph{Appl. Math. Lett.}} \bibinfo{volume}{17}, \bibinfo{number}{10} (\bibinfo{year}{2004}), \bibinfo{pages}{1147--1152}.
\newblock


\bibitem[Liao et~al\mbox{.}(2021)]%
        {Ascend1}
\bibfield{author}{\bibinfo{person}{Heng Liao}, \bibinfo{person}{Jiajin Tu}, \bibinfo{person}{Jing Xia}, \bibinfo{person}{Hu Liu}, \bibinfo{person}{Xiping Zhou}, \bibinfo{person}{Honghui Yuan}, {and} \bibinfo{person}{Yuxing Hu}.} \bibinfo{year}{2021}\natexlab{}.
\newblock \showarticletitle{Ascend: a Scalable and Unified Architecture for Ubiquitous Deep Neural Network Computing : Industry Track Paper}. In \bibinfo{booktitle}{\emph{Proceedings of the {IEEE} International Symposium on High-Performance Computer Architecture, {HPCA}}}. \bibinfo{publisher}{{IEEE}}, \bibinfo{address}{Seoul, South Korea}, \bibinfo{pages}{789--801}.
\newblock


\bibitem[Liao et~al\mbox{.}(2019)]%
        {Ascend2}
\bibfield{author}{\bibinfo{person}{Heng Liao}, \bibinfo{person}{Jiajin Tu}, \bibinfo{person}{Jing Xia}, {and} \bibinfo{person}{Xiping Zhou}.} \bibinfo{year}{2019}\natexlab{}.
\newblock \showarticletitle{DaVinci: {A} Scalable Architecture for Neural Network Computing}. In \bibinfo{booktitle}{\emph{Proceedings of the {IEEE} Hot Chips 31 Symposium (HCS)}}. \bibinfo{publisher}{{IEEE}}, \bibinfo{address}{Cupertino, CA, USA}, \bibinfo{pages}{1--44}.
\newblock


\bibitem[Lin et~al\mbox{.}(2018)]%
        {lcp}
\bibfield{author}{\bibinfo{person}{Xinhan Lin}, \bibinfo{person}{Shouyi Yin}, \bibinfo{person}{Fengbin Tu}, \bibinfo{person}{Leibo Liu}, \bibinfo{person}{Xiangyu Li}, {and} \bibinfo{person}{Shaojun Wei}.} \bibinfo{year}{2018}\natexlab{}.
\newblock \showarticletitle{{LCP:} a layer clusters paralleling mapping method for accelerating inception and residual networks on {FPGA}}. In \bibinfo{booktitle}{\emph{Proceedings of the 55th Annual Design Automation Conference ({DAC})}}. \bibinfo{publisher}{{ACM}}, \bibinfo{address}{San Francisco, CA, USA}, \bibinfo{pages}{16:1--16:6}.
\newblock


\bibitem[Lu et~al\mbox{.}(2017)]%
        {basic-dataflow1}
\bibfield{author}{\bibinfo{person}{Wenyan Lu}, \bibinfo{person}{Guihai Yan}, \bibinfo{person}{Jiajun Li}, \bibinfo{person}{Shijun Gong}, \bibinfo{person}{Yinhe Han}, {and} \bibinfo{person}{Xiaowei Li}.} \bibinfo{year}{2017}\natexlab{}.
\newblock \showarticletitle{FlexFlow: {A} Flexible Dataflow Accelerator Architecture for Convolutional Neural Networks}. In \bibinfo{booktitle}{\emph{Proceedings of the {IEEE} International Symposium on High Performance Computer Architecture ({HPCA})}}. \bibinfo{publisher}{{IEEE} Computer Society}, \bibinfo{address}{Austin, TX, USA}, \bibinfo{pages}{553--564}.
\newblock


\bibitem[Ma et~al\mbox{.}(2017)]%
        {dataflow2}
\bibfield{author}{\bibinfo{person}{Yufei Ma}, \bibinfo{person}{Yu Cao}, \bibinfo{person}{Sarma B.~K. Vrudhula}, {and} \bibinfo{person}{Jae{-}sun Seo}.} \bibinfo{year}{2017}\natexlab{}.
\newblock \showarticletitle{Optimizing Loop Operation and Dataflow in {FPGA} Acceleration of Deep Convolutional Neural Networks}. In \bibinfo{booktitle}{\emph{Proceedings of the {ACM/SIGDA} International Symposium on Field-Programmable Gate Arrays (FPGA)}}. \bibinfo{publisher}{{ACM}}, \bibinfo{address}{Monterey, CA, USA}, \bibinfo{pages}{45--54}.
\newblock


\bibitem[Minsky and Papert(1987)]%
        {slp2}
\bibfield{author}{\bibinfo{person}{Marvin Minsky} {and} \bibinfo{person}{Seymour Papert}.} \bibinfo{year}{1987}\natexlab{}.
\newblock \bibinfo{booktitle}{\emph{Perceptrons - an introduction to computational geometry}}.
\newblock \bibinfo{publisher}{{MIT} Press}, \bibinfo{address}{{}}.
\newblock


\bibitem[Moons et~al\mbox{.}(2017)]%
        {envision}
\bibfield{author}{\bibinfo{person}{Bert Moons}, \bibinfo{person}{Roel Uytterhoeven}, \bibinfo{person}{Wim Dehaene}, {and} \bibinfo{person}{Marian Verhelst}.} \bibinfo{year}{2017}\natexlab{}.
\newblock \showarticletitle{Envision: {A} 0.26-to-10TOPS/W subword-parallel dynamic-voltage-accuracy-frequency-scalable Convolutional Neural Network processor in 28nm {FDSOI}}. In \bibinfo{booktitle}{\emph{Proceedings of the {IEEE} International Solid-State Circuits Conference ({ISSCC})}}. \bibinfo{publisher}{{IEEE}}, \bibinfo{address}{San Francisco, CA, USA}, \bibinfo{pages}{246--247}.
\newblock


\bibitem[Mullapudi et~al\mbox{.}(2016)]%
        {Halide}
\bibfield{author}{\bibinfo{person}{Ravi~Teja Mullapudi}, \bibinfo{person}{Andrew Adams}, \bibinfo{person}{Dillon Sharlet}, \bibinfo{person}{Jonathan Ragan{-}Kelley}, {and} \bibinfo{person}{Kayvon Fatahalian}.} \bibinfo{year}{2016}\natexlab{}.
\newblock \showarticletitle{Automatically scheduling halide image processing pipelines}.
\newblock \bibinfo{journal}{\emph{{ACM} Trans. Graph.}} \bibinfo{volume}{35}, \bibinfo{number}{4} (\bibinfo{year}{2016}), \bibinfo{pages}{83:1--83:11}.
\newblock


\bibitem[Norrie et~al\mbox{.}(2020)]%
        {TPUv2}
\bibfield{author}{\bibinfo{person}{Thomas Norrie}, \bibinfo{person}{Nishant Patil}, \bibinfo{person}{Doe~Hyun Yoon}, \bibinfo{person}{George Kurian}, \bibinfo{person}{Sheng Li}, \bibinfo{person}{James Laudon}, \bibinfo{person}{Cliff Young}, \bibinfo{person}{Norman~P. Jouppi}, {and} \bibinfo{person}{David~A. Patterson}.} \bibinfo{year}{2020}\natexlab{}.
\newblock \showarticletitle{Google's Training Chips Revealed: TPUv2 and TPUv3}. In \bibinfo{booktitle}{\emph{Proceedings of the {IEEE} Hot Chips 32 Symposium ({HCS})}}. \bibinfo{publisher}{{IEEE}}, \bibinfo{address}{Palo Alto, CA, USA}, \bibinfo{pages}{1--70}.
\newblock


\bibitem[NVIDIA(2018)]%
        {nvdla}
\bibfield{author}{\bibinfo{person}{NVIDIA}.} \bibinfo{year}{2018}\natexlab{}.
\newblock \showarticletitle{THE {NVIDIA} DEEP LEARNING ACCELERATOR}. In \bibinfo{booktitle}{\emph{Proceedings of the {IEEE} Hot Chips 30 Symposium (HCS)}}. \bibinfo{publisher}{{IEEE}}, \bibinfo{address}{Cupertino, CA, USA}, \bibinfo{pages}{1--18}.
\newblock


\bibitem[Parashar et~al\mbox{.}(2019)]%
        {timeloop}
\bibfield{author}{\bibinfo{person}{Angshuman Parashar}, \bibinfo{person}{Priyanka Raina}, \bibinfo{person}{Yakun~Sophia Shao}, \bibinfo{person}{Yu{-}Hsin Chen}, \bibinfo{person}{Victor~A. Ying}, \bibinfo{person}{Anurag Mukkara}, \bibinfo{person}{Rangharajan Venkatesan}, \bibinfo{person}{Brucek Khailany}, \bibinfo{person}{Stephen~W. Keckler}, {and} \bibinfo{person}{Joel~S. Emer}.} \bibinfo{year}{2019}\natexlab{}.
\newblock \showarticletitle{Timeloop: {A} Systematic Approach to {DNN} Accelerator Evaluation}. In \bibinfo{booktitle}{\emph{Proceedings of the {IEEE} International Symposium on Performance Analysis of Systems and Software ({ISPASS})}}. \bibinfo{publisher}{{IEEE}}, \bibinfo{address}{Madison, WI, USA}, \bibinfo{pages}{304--315}.
\newblock


\bibitem[Parashar et~al\mbox{.}(2017)]%
        {scnn}
\bibfield{author}{\bibinfo{person}{Angshuman Parashar}, \bibinfo{person}{Minsoo Rhu}, \bibinfo{person}{Anurag Mukkara}, \bibinfo{person}{Antonio Puglielli}, \bibinfo{person}{Rangharajan Venkatesan}, \bibinfo{person}{Brucek Khailany}, \bibinfo{person}{Joel~S. Emer}, \bibinfo{person}{Stephen~W. Keckler}, {and} \bibinfo{person}{William~J. Dally}.} \bibinfo{year}{2017}\natexlab{}.
\newblock \showarticletitle{{SCNN:} An Accelerator for Compressed-sparse Convolutional Neural Networks}. In \bibinfo{booktitle}{\emph{Proceedings of the 44th Annual International Symposium on Computer Architecture ({ISCA})}}. \bibinfo{publisher}{{ACM}}, \bibinfo{address}{Toronto, ON, Canada}, \bibinfo{pages}{27--40}.
\newblock


\bibitem[Radford and Narasimhan(2018)]%
        {GPT}
\bibfield{author}{\bibinfo{person}{Alec Radford} {and} \bibinfo{person}{Karthik Narasimhan}.} \bibinfo{year}{2018}\natexlab{}.
\newblock \showarticletitle{Improving Language Understanding by Generative Pre-Training}. In \bibinfo{booktitle}{\emph{Preprint}}. \bibinfo{publisher}{OpenAI}, \bibinfo{address}{{}}, \bibinfo{pages}{1--12}.
\newblock


\bibitem[Real et~al\mbox{.}(2019)]%
        {nas1}
\bibfield{author}{\bibinfo{person}{Esteban Real}, \bibinfo{person}{Alok Aggarwal}, \bibinfo{person}{Yanping Huang}, {and} \bibinfo{person}{Quoc~V. Le}.} \bibinfo{year}{2019}\natexlab{}.
\newblock \showarticletitle{Regularized Evolution for Image Classifier Architecture Search}. In \bibinfo{booktitle}{\emph{Proceedings of the 33rd Conference on Artificial Intelligence ({AAAI})}}. \bibinfo{publisher}{{AAAI} Press}, \bibinfo{address}{Honolulu, Hawaii, USA}, \bibinfo{pages}{4780--4789}.
\newblock


\bibitem[Rosenblatt(1957)]%
        {slp1}
\bibfield{author}{\bibinfo{person}{Frank Rosenblatt}.} \bibinfo{year}{1957}\natexlab{}.
\newblock \bibinfo{booktitle}{\emph{The perceptron, a perceiving and recognizing automaton Project Para}}.
\newblock \bibinfo{publisher}{Cornell Aeronautical Laboratory}, \bibinfo{address}{{}}.
\newblock


\bibitem[Sandler et~al\mbox{.}(2018)]%
        {mobilenet}
\bibfield{author}{\bibinfo{person}{Mark Sandler}, \bibinfo{person}{Andrew~G. Howard}, \bibinfo{person}{Menglong Zhu}, \bibinfo{person}{Andrey Zhmoginov}, {and} \bibinfo{person}{Liang{-}Chieh Chen}.} \bibinfo{year}{2018}\natexlab{}.
\newblock \showarticletitle{MobileNetV2: Inverted Residuals and Linear Bottlenecks}. In \bibinfo{booktitle}{\emph{Proceedings of the {IEEE} Conference on Computer Vision and Pattern Recognition ({CVPR})}}. \bibinfo{publisher}{Computer Vision Foundation / {IEEE} Computer Society}, \bibinfo{address}{Salt Lake City, UT, USA}, \bibinfo{pages}{4510--4520}.
\newblock


\bibitem[Shao et~al\mbox{.}(2019)]%
        {simba}
\bibfield{author}{\bibinfo{person}{Yakun~Sophia Shao}, \bibinfo{person}{Jason Clemons}, \bibinfo{person}{Rangharajan Venkatesan}, \bibinfo{person}{Brian Zimmer}, \bibinfo{person}{Matthew Fojtik}, \bibinfo{person}{Nan Jiang}, \bibinfo{person}{Ben Keller}, \bibinfo{person}{Alicia Klinefelter}, \bibinfo{person}{Nathaniel Pinckney}, \bibinfo{person}{Priyanka Raina}, \bibinfo{person}{Stephen~G. Tell}, \bibinfo{person}{Yanqing Zhang}, \bibinfo{person}{William~J. Dally}, \bibinfo{person}{Joel Emer}, \bibinfo{person}{C.~Thomas Gray}, \bibinfo{person}{Brucek Khailany}, {and} \bibinfo{person}{Stephen~W. Keckler}.} \bibinfo{year}{2019}\natexlab{}.
\newblock \showarticletitle{Simba: Scaling Deep-Learning Inference with Multi-Chip-Module-Based Architecture}. In \bibinfo{booktitle}{\emph{Proceedings of the {IEEE/ACM} International Symposium on Microarchitecture ({MICRO})}}. \bibinfo{publisher}{{ACM}}, \bibinfo{address}{Columbus, OH, USA}, \bibinfo{pages}{14–27}.
\newblock


\bibitem[Simonyan and Zisserman(2015)]%
        {vgg}
\bibfield{author}{\bibinfo{person}{Karen Simonyan} {and} \bibinfo{person}{Andrew Zisserman}.} \bibinfo{year}{2015}\natexlab{}.
\newblock \showarticletitle{Very Deep Convolutional Networks for Large-Scale Image Recognition}. In \bibinfo{booktitle}{\emph{Proceedings of the International Conference on Learning Representations ({ICLR})}}. \bibinfo{publisher}{Computational and Biological Learning Society}, \bibinfo{address}{San Diego, CA, USA}, \bibinfo{pages}{1--14}.
\newblock


\bibitem[Song et~al\mbox{.}(2020)]%
        {DRQ}
\bibfield{author}{\bibinfo{person}{Zhuoran Song}, \bibinfo{person}{Bangqi Fu}, \bibinfo{person}{Feiyang Wu}, \bibinfo{person}{Zhaoming Jiang}, \bibinfo{person}{Li Jiang}, \bibinfo{person}{Naifeng Jing}, {and} \bibinfo{person}{Xiaoyao Liang}.} \bibinfo{year}{2020}\natexlab{}.
\newblock \showarticletitle{{DRQ:} Dynamic Region-based Quantization for Deep Neural Network Acceleration}. In \bibinfo{booktitle}{\emph{Proceedings of the 47th {ACM/IEEE} Annual International Symposium on Computer Architecture ({ISCA})}}. \bibinfo{publisher}{{IEEE}}, \bibinfo{address}{Valencia, Spain}, \bibinfo{pages}{1010--1021}.
\newblock


\bibitem[Szegedy et~al\mbox{.}(2015)]%
        {inception}
\bibfield{author}{\bibinfo{person}{Christian Szegedy}, \bibinfo{person}{Wei Liu}, \bibinfo{person}{Yangqing Jia}, \bibinfo{person}{Pierre Sermanet}, \bibinfo{person}{Scott~E. Reed}, \bibinfo{person}{Dragomir Anguelov}, \bibinfo{person}{Dumitru Erhan}, \bibinfo{person}{Vincent Vanhoucke}, {and} \bibinfo{person}{Andrew Rabinovich}.} \bibinfo{year}{2015}\natexlab{}.
\newblock \showarticletitle{Going deeper with convolutions}. In \bibinfo{booktitle}{\emph{Proceedings of the {IEEE} Conference on Computer Vision and Pattern Recognition ({CVPR})}}. \bibinfo{publisher}{{IEEE} Computer Society}, \bibinfo{address}{Boston, MA, USA}, \bibinfo{pages}{1--9}.
\newblock


\bibitem[Talpes et~al\mbox{.}(2022)]%
        {dojo}
\bibfield{author}{\bibinfo{person}{Emil Talpes}, \bibinfo{person}{Douglas Williams}, {and} \bibinfo{person}{Debjit~Das Sarma}.} \bibinfo{year}{2022}\natexlab{}.
\newblock \showarticletitle{{DOJO:} The Microarchitecture of Tesla's Exa-Scale Computer}. In \bibinfo{booktitle}{\emph{Proceedings of the {IEEE} Hot Chips 34 Symposium (HCS)}}. \bibinfo{publisher}{{IEEE}}, \bibinfo{address}{Cupertino, CA, USA}, \bibinfo{pages}{1--28}.
\newblock


\bibitem[Tan et~al\mbox{.}(2021)]%
        {nnbaton}
\bibfield{author}{\bibinfo{person}{Zhanhong Tan}, \bibinfo{person}{Hongyu Cai}, \bibinfo{person}{Runpei Dong}, {and} \bibinfo{person}{Kaisheng Ma}.} \bibinfo{year}{2021}\natexlab{}.
\newblock \showarticletitle{{NN}-Baton: {DNN} Workload Orchestration and Chiplet Granularity Exploration for Multichip Accelerators}. In \bibinfo{booktitle}{\emph{Proceedings of the {IEEE} Annual International Symposium on Computer Architecture ({ISCA})}}. \bibinfo{publisher}{{IEEE}}, \bibinfo{address}{Valencia, Spain}, \bibinfo{pages}{1013--1026}.
\newblock


\bibitem[Tarnawski et~al\mbox{.}(2020)]%
        {dnn-partitioning}
\bibfield{author}{\bibinfo{person}{Jakub Tarnawski}, \bibinfo{person}{Amar Phanishayee}, \bibinfo{person}{Nikhil~R. Devanur}, \bibinfo{person}{Divya Mahajan}, {and} \bibinfo{person}{Fanny~Nina Paravecino}.} \bibinfo{year}{2020}\natexlab{}.
\newblock \showarticletitle{Efficient Algorithms for Device Placement of {DNN} Graph Operators}. In \bibinfo{booktitle}{\emph{Advances in Neural Information Processing Systems ({NeurIPS})}}, \bibfield{editor}{\bibinfo{person}{Hugo Larochelle}, \bibinfo{person}{Marc'Aurelio Ranzato}, \bibinfo{person}{Raia Hadsell}, \bibinfo{person}{Maria{-}Florina Balcan}, {and} \bibinfo{person}{Hsuan{-}Tien Lin}} (Eds.). \bibinfo{publisher}{OpenReview.net}, \bibinfo{address}{Virtual}, \bibinfo{pages}{1--13}.
\newblock


\bibitem[Tenstorrent(2021)]%
        {Grayskull1}
\bibfield{author}{\bibinfo{person}{Tenstorrent}.} \bibinfo{year}{2021}\natexlab{}.
\newblock \bibinfo{title}{Grayskull}.
\newblock \bibinfo{howpublished}{\url{https://tenstorrent.com/grayskull/}}.
\newblock


\bibitem[Vaswani et~al\mbox{.}(2017)]%
        {transformer}
\bibfield{author}{\bibinfo{person}{Ashish Vaswani}, \bibinfo{person}{Noam Shazeer}, \bibinfo{person}{Niki Parmar}, \bibinfo{person}{Jakob Uszkoreit}, \bibinfo{person}{Llion Jones}, \bibinfo{person}{Aidan~N. Gomez}, \bibinfo{person}{Lukasz Kaiser}, {and} \bibinfo{person}{Illia Polosukhin}.} \bibinfo{year}{2017}\natexlab{}.
\newblock \showarticletitle{Attention is All you Need}. In \bibinfo{booktitle}{\emph{Advances in Neural Information Processing Systems {(NIPS)}}}, \bibfield{editor}{\bibinfo{person}{Isabelle Guyon}, \bibinfo{person}{Ulrike von Luxburg}, \bibinfo{person}{Samy Bengio}, \bibinfo{person}{Hanna~M. Wallach}, \bibinfo{person}{Rob Fergus}, \bibinfo{person}{S.~V.~N. Vishwanathan}, {and} \bibinfo{person}{Roman Garnett}} (Eds.). \bibinfo{publisher}{OpenReview.net}, \bibinfo{address}{Long Beach, CA, USA}, \bibinfo{pages}{5998--6008}.
\newblock


\bibitem[Wechsler et~al\mbox{.}(2019)]%
        {NNP-I}
\bibfield{author}{\bibinfo{person}{Ofri Wechsler}, \bibinfo{person}{Michael Behar}, {and} \bibinfo{person}{Bharat Daga}.} \bibinfo{year}{2019}\natexlab{}.
\newblock \showarticletitle{Spring Hill {(NNP-I} 1000) Intel's Data Center Inference Chip}. In \bibinfo{booktitle}{\emph{Proceedings of the {IEEE} Hot Chips 31 Symposium ({HCS})}}. \bibinfo{publisher}{{IEEE}}, \bibinfo{address}{Cupertino, CA, USA}, \bibinfo{pages}{1--12}.
\newblock


\bibitem[Weng et~al\mbox{.}(2020)]%
        {dse}
\bibfield{author}{\bibinfo{person}{Jian Weng}, \bibinfo{person}{Sihao Liu}, \bibinfo{person}{Vidushi Dadu}, \bibinfo{person}{Zhengrong Wang}, \bibinfo{person}{Preyas Shah}, {and} \bibinfo{person}{Tony Nowatzki}.} \bibinfo{year}{2020}\natexlab{}.
\newblock \showarticletitle{{DSAGEN:} Synthesizing Programmable Spatial Accelerators}. In \bibinfo{booktitle}{\emph{Proceedings of the 47th {ACM/IEEE} Annual International Symposium on Computer Architecture ({ISCA})}}. \bibinfo{publisher}{{IEEE}}, \bibinfo{address}{Valencia, Spain}, \bibinfo{pages}{268--281}.
\newblock


\bibitem[Xiao et~al\mbox{.}(2021)]%
        {hasco}
\bibfield{author}{\bibinfo{person}{Qingcheng Xiao}, \bibinfo{person}{Size Zheng}, \bibinfo{person}{Bingzhe Wu}, \bibinfo{person}{Pengcheng Xu}, \bibinfo{person}{Xuehai Qian}, {and} \bibinfo{person}{Yun Liang}.} \bibinfo{year}{2021}\natexlab{}.
\newblock \showarticletitle{{HASCO:} Towards Agile HArdware and Software CO-design for Tensor Computation}. In \bibinfo{booktitle}{\emph{Proceedings of the 48th {ACM/IEEE} Annual International Symposium on Computer Architecture ({ISCA})}}. \bibinfo{publisher}{{IEEE}}, \bibinfo{address}{Valencia, Spain}, \bibinfo{pages}{1055--1068}.
\newblock


\bibitem[Xie et~al\mbox{.}(2019)]%
        {random-gen}
\bibfield{author}{\bibinfo{person}{Saining Xie}, \bibinfo{person}{Alexander Kirillov}, \bibinfo{person}{Ross~B. Girshick}, {and} \bibinfo{person}{Kaiming He}.} \bibinfo{year}{2019}\natexlab{}.
\newblock \showarticletitle{Exploring Randomly Wired Neural Networks for Image Recognition}. In \bibinfo{booktitle}{\emph{Proceedings of the {IEEE/CVF} International Conference on Computer Vision ({ICCV})}}. \bibinfo{publisher}{{IEEE}}, \bibinfo{address}{Seoul, South Korea}, \bibinfo{pages}{1284--1293}.
\newblock


\bibitem[Yang(2019)]%
        {NNP-T}
\bibfield{author}{\bibinfo{person}{Andrew Yang}.} \bibinfo{year}{2019}\natexlab{}.
\newblock \showarticletitle{Deep Learning Training At Scale Spring Crest Deep Learning Accelerator (Intel{\textregistered} Nervana{\texttrademark} {NNP-T)}}. In \bibinfo{booktitle}{\emph{Proceedings of the {IEEE} Hot Chips 31 Symposium (HCS)}}. \bibinfo{publisher}{{IEEE}}, \bibinfo{address}{Cupertino, CA, USA}, \bibinfo{pages}{1--20}.
\newblock


\bibitem[Yang et~al\mbox{.}(2020)]%
        {interstellar}
\bibfield{author}{\bibinfo{person}{Xuan Yang}, \bibinfo{person}{Mingyu Gao}, \bibinfo{person}{Qiaoyi Liu}, \bibinfo{person}{Jeff Setter}, \bibinfo{person}{Jing Pu}, \bibinfo{person}{Ankita Nayak}, \bibinfo{person}{Steven Bell}, \bibinfo{person}{Kaidi Cao}, \bibinfo{person}{Heonjae Ha}, \bibinfo{person}{Priyanka Raina}, \bibinfo{person}{Christos Kozyrakis}, {and} \bibinfo{person}{Mark Horowitz}.} \bibinfo{year}{2020}\natexlab{}.
\newblock \showarticletitle{Interstellar: Using Halide's Scheduling Language to Analyze DNN Accelerators}. In \bibinfo{booktitle}{\emph{Proceedings of the International Conference on Architectural Support for Programming Languages and Operating Systems ({ASPLOS})}}. \bibinfo{publisher}{{ACM}}, \bibinfo{address}{Lausanne, Switzerland}, \bibinfo{pages}{369–383}.
\newblock


\bibitem[Zheng et~al\mbox{.}(2022a)]%
        {amos}
\bibfield{author}{\bibinfo{person}{Size Zheng}, \bibinfo{person}{Renze Chen}, \bibinfo{person}{Anjiang Wei}, \bibinfo{person}{Yicheng Jin}, \bibinfo{person}{Qin Han}, \bibinfo{person}{Liqiang Lu}, \bibinfo{person}{Bingyang Wu}, \bibinfo{person}{Xiuhong Li}, \bibinfo{person}{Shengen Yan}, {and} \bibinfo{person}{Yun Liang}.} \bibinfo{year}{2022}\natexlab{a}.
\newblock \showarticletitle{{AMOS:} enabling automatic mapping for tensor computations on spatial accelerators with hardware abstraction}. In \bibinfo{booktitle}{\emph{Proceedings of the 49th Annual International Symposium on Computer Architecture {(ISCA)}}}. \bibinfo{publisher}{{ACM}}, \bibinfo{address}{New York, New York, USA}, \bibinfo{pages}{874--887}.
\newblock


\bibitem[Zheng et~al\mbox{.}(2022b)]%
        {atomic}
\bibfield{author}{\bibinfo{person}{Shixuan Zheng}, \bibinfo{person}{Xianjue Zhang}, \bibinfo{person}{Leibo Liu}, \bibinfo{person}{Shaojun Wei}, {and} \bibinfo{person}{Shouyi Yin}.} \bibinfo{year}{2022}\natexlab{b}.
\newblock \showarticletitle{Atomic Dataflow based Graph-Level Workload Orchestration for Scalable {DNN} Accelerators}. In \bibinfo{booktitle}{\emph{Proceedings of the {IEEE} International Symposium on High-Performance Computer Architecture ({HPCA})}}. \bibinfo{publisher}{{IEEE}}, \bibinfo{address}{Seoul, South Korea}, \bibinfo{pages}{475--489}.
\newblock


\bibitem[Zheng et~al\mbox{.}(2020)]%
        {irregular}
\bibfield{author}{\bibinfo{person}{Shixuan Zheng}, \bibinfo{person}{Xianjue Zhang}, \bibinfo{person}{Daoli Ou}, \bibinfo{person}{Shibin Tang}, \bibinfo{person}{Leibo Liu}, \bibinfo{person}{Shaojun Wei}, {and} \bibinfo{person}{Shouyi Yin}.} \bibinfo{year}{2020}\natexlab{}.
\newblock \showarticletitle{Efficient Scheduling of Irregular Network Structures on {CNN} Accelerators}.
\newblock \bibinfo{journal}{\emph{{IEEE} Transactions on Computer-Aided Design of Integrated Circuits and Systems ({TCAD})}} \bibinfo{volume}{39}, \bibinfo{number}{11} (\bibinfo{year}{2020}), \bibinfo{pages}{3408--3419}.
\newblock


\bibitem[Zimmer et~al\mbox{.}(2019)]%
        {simba-vlsi}
\bibfield{author}{\bibinfo{person}{Brian Zimmer}, \bibinfo{person}{Rangharajan Venkatesan}, \bibinfo{person}{Yakun~Sophia Shao}, \bibinfo{person}{Jason Clemons}, \bibinfo{person}{Matthew Fojtik}, \bibinfo{person}{Nan Jiang}, \bibinfo{person}{Ben Keller}, \bibinfo{person}{Alicia Klinefelter}, \bibinfo{person}{Nathaniel~Ross Pinckney}, \bibinfo{person}{Priyanka Raina}, \bibinfo{person}{Stephen~G. Tell}, \bibinfo{person}{Yanqing Zhang}, \bibinfo{person}{William~J. Dally}, \bibinfo{person}{Joel~S. Emer}, \bibinfo{person}{C.~Thomas Gray}, \bibinfo{person}{Stephen~W. Keckler}, {and} \bibinfo{person}{Brucek Khailany}.} \bibinfo{year}{2019}\natexlab{}.
\newblock \showarticletitle{A 0.11 pJ/Op, 0.32-128 TOPS, Scalable Multi-Chip-Module-based Deep Neural Network Accelerator with Ground-Reference Signaling in 16nm}. In \bibinfo{booktitle}{\emph{Proceedings of the {IEEE} Symposium on VLSI Circuits ({VLSI})}}. \bibinfo{publisher}{{IEEE}}, \bibinfo{address}{Kyoto, Japan}, \bibinfo{pages}{300}.
\newblock


\bibitem[Zoph et~al\mbox{.}(2018)]%
        {nasnet}
\bibfield{author}{\bibinfo{person}{Barret Zoph}, \bibinfo{person}{Vijay Vasudevan}, \bibinfo{person}{Jonathon Shlens}, {and} \bibinfo{person}{Quoc~V. Le}.} \bibinfo{year}{2018}\natexlab{}.
\newblock \showarticletitle{Learning Transferable Architectures for Scalable Image Recognition}. In \bibinfo{booktitle}{\emph{{IEEE} Conference on Computer Vision and Pattern Recognition, ({CVPR})}}. \bibinfo{publisher}{Computer Vision Foundation / {IEEE} Computer Society}, \bibinfo{address}{Salt Lake City, UT, USA}, \bibinfo{pages}{8697--8710}.
\newblock


\end{thebibliography}

%%
%% If your work has an appendix, this is the place to put it.
% \appendix

\end{document}